\documentclass{emulateapj}

\usepackage{graphics,graphicx,hyperref}
\usepackage{epstopdf}
\usepackage{amssymb, amsmath, amsthm}
\usepackage[version=3]{mhchem}
\usepackage{graphicx}
\usepackage{multirow}
\usepackage{booktabs,threeparttable}
\usepackage{subfigure}
\usepackage{xr} 
\usepackage[section]{placeins}
\usepackage{amsmath}
\usepackage{wrapfig}
\usepackage{wrapfig}
\usepackage{tabularx}
\usepackage{float}
\usepackage{indentfirst}
\usepackage{sidecap}
\usepackage{subfigure}
\usepackage{color} 
\usepackage[normalem]{ulem}

\usepackage[authordate,backend=biber,maxnames=1,minnames=1]{}

\begin{document}

\title{Systematic analysis of spectral energy distributions and the dust opacity indices for Class 0 young stellar objects}

\author{Jennifer I-Hsiu Li\altaffilmark{1,2}, Hauyu Baobab Liu\altaffilmark{3}, Yasuhiro Hasegawa\altaffilmark{4}, Naomi Hirano\altaffilmark{1}}
\affil{$^1$Institute of Astronomy and Astrophysics, Academia Sinica, Taipei, Taiwan}
\affil{$^2$Department of Astronomy, University of Illinois at Urbana-Champaign, Urbana, IL, USA}
\affil{$^3$European Southern Observatory (ESO), Garching, Germany}
\affil{$^4$Jet Propulsion Laboratory, California Institute of Technology, Pasadena, CA, USA}

\begin{abstract}
We are motivated by the recent measurements of dust opacity indices ($\beta$) around young stellar objects (YSOs), which suggest that efficient grain growth may have occurred earlier than the Class I stage.
The present work makes use of abundant archival interferometric observations at submillimeter,
millimeter, and centimeter wavelength bands to examine grain growth signatures in the dense inner regions ($<$1000 AU) of nine Class 0 YSOs.
A systematic data analysis is performed to derive dust temperatures,  optical depths, and dust opacity indices based on single-component modified black body fittings to the spectral energy distributions (SEDs).
The fitted dust opacity indices ($\beta$) are in a wide range of 0.3 to 2.0 when single-component SED fitting is adopted.
Four out of the nine observed sources show $\beta$ lower than 1.7, the typical value of the interstellar dust. 
Low dust opacity index (or spectral index) values may be explained by the effect of dust grain growth, which makes $\beta<$1.7.
Alternatively, the very small observed values of $\beta$ may be interpreted by the presence of deeply embedded hot inner disks, which only significantly contribute to the observed fluxes at long wavelength bands.
This possibility can be tested by the higher angular resolution imaging observations of ALMA, or more detailed sampling of SEDs in the millimeter and centimeter bands.
The $\beta$ values of the remaining five sources are close to or consistent with 1.7, indicating that grain growth would start to significantly reduce the values of $\beta$ no earlier than the late-Class 0 stage for these YSOs.
\end{abstract}


\section{Introduction} \label{sec:intro}

When do the Keplerian rotating protoplanetary disks form? When and how does the maximum dust grain size grow from micron-size to millimeter-size in the protoplanetary disks? 
These are crucial questions when addressing the formation of planetesimals and planetary systems in star-forming regions.
Observationally, the maximum size of dust grains may be constrained by the observations of spectral energy distributions (SEDs) and the dust opacity indices $\beta$ \citep{draine2006submillimeter}. 
The typical $\beta$ value for dust in the optically thin interstellar medium (ISM) is $\sim$1.7 (e.g., \citealt{Lin2016}); grain growth up to millimeter sizes will result in $\beta<1$ \citep[e.g.,][]{beckwith1991, draine2006submillimeter, ricci2010b, Kataoka2014}.
Previous observations have shown that some Class I and II sources have $\beta$ values smaller than that of ISM \citep[e.g.,][]{beckwith1991,ricci2010a,ricci2010b,miotello2014}.
On the other hand, \cite{forbrich2015} studied the starless core FeSt 1-457 in the pipe nebula and found that $\beta$ in the inner core region shows no significant difference from local clouds, which is similar to the values associated with  ISM. 
Therefore, studying Class 0 YSOs may provide some clues to the starting point when grain growth begins to significantly reduce the value of $\beta$.
A comparison of sources at various evolutionary stages will further constrain the timescale of this process.

Derivations of $\beta$ from deeply embedded Class 0 YSOs with observations merely at shorter than $\sim$1 mm wavelengths can be degenerated due to the high optical depth.
\citet{Jorgensen2009} have performed a study of 20 Class 0 and I YSOs with the 0.88 and 1.3 mm observations of SMA, and derived spectral indices $\alpha$ around 2.4. 
Their interpretation was either that significant grain growth has occurred in the observed sources with $\beta$ lower than 2.0, or the dust emission is optically thick and the Rayleigh-Jeans approximation breaks down.
Based on CARMA observations, \cite{Kwon2009} and \cite{chiang2012envelope} found $\beta$$\sim$1 in L1448 IRS2, L1448 IRS3B and L1157 in the envelope ($>10^3$ AU scale). 
They claimed a radial dependence of $\beta$ in these sources, which might suggest that grain growth is quicker in denser regions. 
However, most past studies assumed that Class 0 envelopes are optically thin and fall in the Rayleigh-Jeans regime at observed (sub)millimeter wavelengths, which may potentially result in underestimating $\beta$.

In this work, we utilized the archival data of National Radio Astronomy Observatory (NRAO) Very Large Array (VLA) [now the Karl G. Jansky Very Large Array (JVLA)], the Submillimeter Array (SMA), the Combined Array for Research in Millimeter-wave Astronomy (CARMA), the Nobeyama Radio Observatory (NRO) Nobeyama Millimeter Array (NMA), and the Atacama Large Millimeter Array (ALMA), for nine Class 0 YSOs.
The good sampling of SEDs from the centimeter to the submillimeter bands, incorporated with archival infrared photometry, allows simultaneously fitting the dust temperature, dust opacity and $\beta$.
Observations and our data reduction are described in Section \ref{sec:observation}.
Section \ref{sec:procedure} summarizes our procedure of data analysis.
The results are presented in Section \ref{sec:result}. 
We discuss our results, their implications on grain growth and alternative explanations in Section \ref{sec:discussion}.
Our conclusions are given in Section \ref{sec:conclusion}.

\section{Observations and Data Reduction}\label{sec:observation}

In the following sections, we describe how our target sources were selected based on the spatial resolution and spectral band completeness of the archival observations. 
Also, we briefly summarize the data calibration procedures.
Our targets are listed in Table \ref{tab:source}.
The observations are tabulated in Table \ref{tab:obs}.

\subsection{Target sources}
To simultaneously determine dust temperature, optical depth and $\beta$ using a single component modified black body fitting (Section \ref{sub:sed}), we select Class 0 sources that have archival data at no less than three frequency bands in the millimeter and submillimeter wavelength range.
In addition, we require the selected sources to present at least one significant detection at $>$3 mm wavelengths.
This is because the dust emission tends to be optically thinner at these wavelengths, so that we can obtain a better constrain on the value of $\beta$.
Limited by the (low) angular resolution of the archival SMA observations (Section \ref{sub:obssma}), we only consider nearby targets that are located within 420 pc from the Sun for better spatial resolution.
\begin{table*}
\caption{List of our targets}
\centering  
\label{tab:source} 
\begin{tabular}{lllll}
\toprule
Source Name & R.A. and Dec. & Region & Distance (pc) & Class   \\
\midrule
L1448CN 			&03:25:38.80, +30:45:05.0& Perseus& 240 & 0\\[2pt]
NGC1333 IRAS2A           & 03:28:55.70, +31:14:37.0 & Perseus  & 240 & 0   \\[2pt]
NGC1333 IRAS4 (A1, A2)    & 03:29:10.50, +31:13:31.0 & Perseus  & 240 & 0   \\[2pt]
Barnard 1b (N, S)            & 03:33:21.14, +31:07:35.3 & Perseus  & 240 & 0   \\[2pt]
L1527                   & 04:39:53.90, +26:03:10.0 & Taurus   & 140 & 0/I \\[2pt]
Serpens FIRS 1          & 18:29:49.80, +01:15:20.6 & Serpens  & 415 &  0/I  \\[2pt]
L1157mm                 & 20:39:06.19, +68:02:15.9 & Isolated & 325 & 0   \\[2pt]
\bottomrule
\end{tabular}
\vspace{0.3cm}
\end{table*}
Based on these criteria, nine well-studied class 0 YSOs are selected and investigated. 
We quote their distances from \citet{chen2013SMA}, except for Serpens FIRS1 (also known as Serpens SMM1), for which we adopt the most recent measured distance \citep[see][for further information]{dzib2010vlba}.
Seven out of nine of the sources are classified as Class 0 YSOs, and the remaining two are assumed to be in the transition phase between the Class 0 and Class I stages \citep{chen2013SMA, froebrich2005youngest}. 
Basic properties of our samples are summarized in Table \ref{tab:source}. 
We refer the reader to \citet{chen2013SMA} for other detailed information about these YSOs. 

\begin{table*}
\caption{Summary of Observations}  
\centering
\label{tab:obs}
\begin{tabular}{ccccl}
\toprule
Source  & Frequency   & Telescope &UV Coverage  & Observing Date \\
        & (GHz)        && (k$\lambda$) &      \\
\midrule
L1448CN 
	& 37		& VLA	& 10--1290 & 2013 Oct. 28  \\
	& 43		& VLA	& 10--1300	& 1996 Feb. 23, 24, 25\\
	& 230	& SMA & 3--390 & 2004 Nov. 7 \\
    &	&  &  & 2011 Sep. 12 \\
	&	&  &  & 2012 Jan. 7 \\
	& 340	&SMA & 21--579& 2010 Jul. 19, Sep. 17  \\
\midrule
NGC1333 IRAS 2A	
	& 37		& VLA	& 20--1390	&  2013 Nov. 4  \\
	& 43		& VLA &5--483 & 2004 Mar. 28  \\
    & 100		& CARMA & 5--125 & 2008 Oct. 14, 20\\
	& 194	& SMA &	6--50	& 2008 Aug. 19	 \\
	& 340	& SMA &	10--256	& 2010 Oct. 14	 \\
	&		& 	 &		& 2013 Oct. 21	 \\
\midrule
NGC1333 IRAS4 (A1,A2)   
	& 37		& VLA	& 21--1420	& 2013 Oct. 21  \\
	& 43   	& VLA &   4--500    & 2014 Oct. 13    \\
	& 230   	&  SMA &8--394	& 2004 Nov. 22  \\
		&		& 	 &		& 2006 Jan. 17	 \\
		&		& 	 &		& 2011 Sep. 12	 \\
	& 340   	&  SMA &10--530	& 2004 Dec. 05, 06  \\
		&		& 	 &		& 2009 Jan. 27, Feb. 24	 \\
		&		& 	 &		& 2011 Jan. 24, 25	 \\
\midrule
Barbard 1B (N, S) 
	& 37		& VLA	& 10--1290	 & 2013 Nov. 7 \\
    & 100		& NMA	& 0--100 & 1998 Jan. -- May \\
	& 230	&  SMA &5--170	 &	2013 Nov. 19, 23, Dec. 6	\\
	&		& 	 &		& 2014 Jan. 11 	 \\
    & 280	& SMA	& 3--75 &	2008 Sep. 03\\
	& 340	& SMA &12--88	 &	2012 Nov. 10	\\
\midrule
L1527    
	& 43 & VLA &10--2000 & 1996 Feb. 23, 24, 25	 \\
	&&	&		& 2004 Feb. 11	 \\
	& 90		& CARMA & 5--560 & 2009 Oct. 11		 \\
	&		&	 	&	& 2010 Dec. 02			 \\
	& 230   	& SMA &4--400   & 2011 Sep. 02, 09            \\
	&		&	 &		& 2012 Jan. 08 			 \\
	& 345	& ALMA &	12--432	&	2012 Aug. 29	 \\
\midrule
Serpens FIRS 1  	
	& 43 		& VLA & 3--426 &2004 Mar. 11 \\
        & 230   	& SMA &37--386	& 2009 Jul. 28  \\
        & 340   	& SMA &6--45  	& 2012 Jul. 01  \\
\midrule
L1157   
	& 43 		& VLA & 3--492 & 2001 Nov. 29  \\
	&		&	&	& 2004 Mar. 14  \\
	&		&	&	& 2004 Nov. 26  \\
	& 230   	& SMA &4--52  	& 2005 Jul. 06  \\
	&       	&   	   &     	& 2009 Sep. 15  \\
        & 340   	& SMA &7--79  	& 2005 Jul. 03, Aug. 22  \\
\bottomrule
\end{tabular}
\vspace{0.3cm}
\end{table*}

\subsection{Submillimeter, millimeter, and radio interferometric data} \label{sec:obs_1}
We retrieve centimeter band observations from VLA, millimeter band observations from VLA, CARMA, NMA and SMA, and submillimeter band observations from SMA and ALMA.
Each set of data is calibrated using their official data reduction software packages.
We use the AIPS \citep{greisen2002aips} software package to reduce the VLA data taken before 2010, the  CASA \citep{mcmullin2007casa} software package to reduce (J)VLA data taken after 2010 and ALMA data, the MIR IDL software package \citep{qi2005mir} for the SMA data reduction, the UVPROC2 software package for the NMA data reduction, and the MIRIAD \citep{sault1995retrospective} software package for the CARMA data reduction.
Finally, MIRIAD is used to evaluate visibility amplitudes as a function of {\it uv} distance. 
When multiple observations are accessible at a certain wavelength for a target source, we combine all the available visibility data in order to maximize the signal-to-noise ratio (SNR). 

Some of our targets are binaries. 
In such cases, we first generate the {\tt clean} component maps using the MIRIAD task CLEAN.
Then we use the MIRIAD task {\tt UVMODEL} for modeling and subtracting the secondary component from the calibrated visibility data. 

A brief summary of each observation and additional treatments in the data reduction is provided below.

\subsubsection{VLA Observations}
We have performed deep $\sim$43 GHz observations towards NGC1333 IRAS4A using JVLA in 2014 (see \cite{Liu2016IRAS4A} for the details).
For all the Perseus sources in our sample, we obtain the public data from the VLA Nascent Disk and Multiplicity (VANDAM) survey \citep{cox2015VANDAM, tobin2014vla}, 
which observed all known protostars in the Perseus molecular cloud in 2013--2014. 
We only include the upper side band (USB, centered at 37 GHz) data for our analysis, owning to the poor data quality in the lower side band (LSB, centered at 29 GHz).
These data taken with the upgraded JVLA have a very high sensitivity and broad bandwidth coverages, and hence allow us to derive good constraints on $\beta$ in SED fitting.
We obtain 43GHz data from the VLA data archive for all the sources except B1b (N, S), which has a low SNR.
These observations were conducted between 1996 and 2004. 

We combined all available array configurations (including A, B, C, BnC and D) with good SNRs. The combined visibility data in general provide visibility data up to few thousands $k\lambda$ {\it uv} distances.
To compare the (J)VLA data with the SMA data, we trim all the data at $\gtrsim$800 $k\lambda$ {\it uv} distances.

\subsubsection{CARMA observations}
The 90 GHz data for L1527 and 100 GHz data for NGC1333 IRAS2A are obtained from the CARMA data archive. 
We combine data taken with the A and C array configurations to match the {\it uv} distance of other observations for L1527.  

\subsubsection{NMA observations}
The 100 GHz data for Barnard 1b-N/S were observed with the NMA. The details of the observations are described in \citet{hirano2014}.

\subsubsection{SMA observations}\label{sub:obssma}
We obtain 183 GHz, 230 GHz, 280 GHz and 340 GHz data from the SMA data archive. 
The antenna configurations include subcompact, compact, extended and very extended. 
The {\it uv} coverage of the observations is typically from a few to few hundred $k\lambda$, which corresponds to ${10}^{3}$-${10}^{5}$AU in physical scale for our targets. 
We use the MIRIAD task {\tt UVLIN} to generate the spectral line-free continuum data.
For the case that {\tt UVLIN} cannot provide a good fit to the continuum level, we use the MIRIAD  {\tt UVAVER} routine to average the continuum data over wide channel ranges.

\subsubsection{ALMA observations}
For L1527, both ALMA and SMA 340 GHz data exist. 
We choose to use the ALMA data only, because of their better sensitivity and angular resolution. 
The observation was conducted in ALMA Cycle 0 in 2012  \citep[see, e.g., ][for the details]{Sakai2014}.
We separate the continuum data using the CASA task {\tt UVCONTSUB}.

\subsection{FIR data}
In order to resolve the degeneracy between $\beta$ and the dust temperature, we include the {\it Spitzer} Multiband Imaging Photometer (MIPS) 24 and 70 $\micron$ data to our SED fitting.
Considering that the typical dust temperature for our targets is about 10-100 K, these short wavelength observations can cover the high frequency wing in SEDs. 
Note that the Spitzer observations are lower in spatial resolution (6$''$ at 24 $\micron$ and 18$''$ at 70$\micron$), compared with radio observations. 
The low resolution of these observations may lead to underestimate temperatures and dust opacity indices.
A quantitative discussion is provided in Section \ref{sec:tempandbeta}.

\section{Flows of Data Analysis Procedures}\label{sec:procedure}
This section introduces the procedure we used for a systematic data analysis. 
It is motivated by a widely adopted geometric picture of YSOs \citep{Shu1987}; YSOs can be modeled as spatially compact disk(-like) structures surrounded by large-scale extended envelopes. 
We first perform an automatized binning for the azimuthally averaged visibility data in the {\it uv} distance domain (Section \ref{sub:binning}). 
We subsequently perform Gaussian decomposition for the binned visibility amplitudes to separate the compact and extended components (Section \ref{sec:data_ana_2}). 
Finally, we conduct the modified black body SED fittings for these components to derive dust temperature, dust opacity, and opacity index (Section \ref{sub:sed}).
A number of error sources are considered in the SED fitting, which are described in detail in Section \ref{sub:freefree}.

\subsection{Visibility Amplitude Analysis: Optimizing Bin Sizes}\label{sub:binning}
We dynamically bin the visibility amplitudes to properly weight the observations at different {\it uv} distance ranges. 
We first compute the azimuthally averaged visibility amplitudes at each {\it uv} distance using the MIRIAD task {\tt UVAMP}. Once the visibility amplitudes are obtained, optimal bin sizes are searched 
such that the following two criteria will be fulfilled: (1) low SNR data points ($<$0.8 of the median value of unbinned data points) are binned with adjacent data points, and (2) long baseline data points with low SNR are binned with wider bins under the condition that the maximum bin width is restricted to the empirically assigned upper limit:
\begin{equation} 
\frac{0.1\times \mbox{uv-distance}{}^{2}}{max(\mbox{uv-distance})}. 
\end{equation}
The former is to ensure good SNRs, while the latter is to avoid over smearing the spatial information.

\subsection{Gaussian Decomposition} \label{sec:data_ana_2}
The typical sizes of circumstellar envelopes and (proto-)disks of Class 0 YSO are of a few thousands AU and $<$100 AU \citep{williams2011}, respectively. 
Most of our retrieved radio data cover {\it uv} distances of from a few to few hundreds k$\lambda$, which corresponds to ${10}^{5}$-${10}^{2}$AU in physical scale at the distance of  $\sim$ 300 pc.  Therefore, these observations are, in most cases, insufficient to resolve, or at most resolve marginally the hypothesized compact disk-like component.
On the other hand, they are good enough to resolve the extended circumstellar envelops.
Nevertheless, the assumed compact and extended components should have very different visibility amplitude distributions, and hence would be possible to systematically decompose the visibility data into these two components.
We have inspected all the data of our targets and confirmed that both the compact and the extended components are present clearly in their visibility amplitude distributions.

\begin{figure}[t]
\hspace{-0.3cm}
\includegraphics[width=9.5cm]{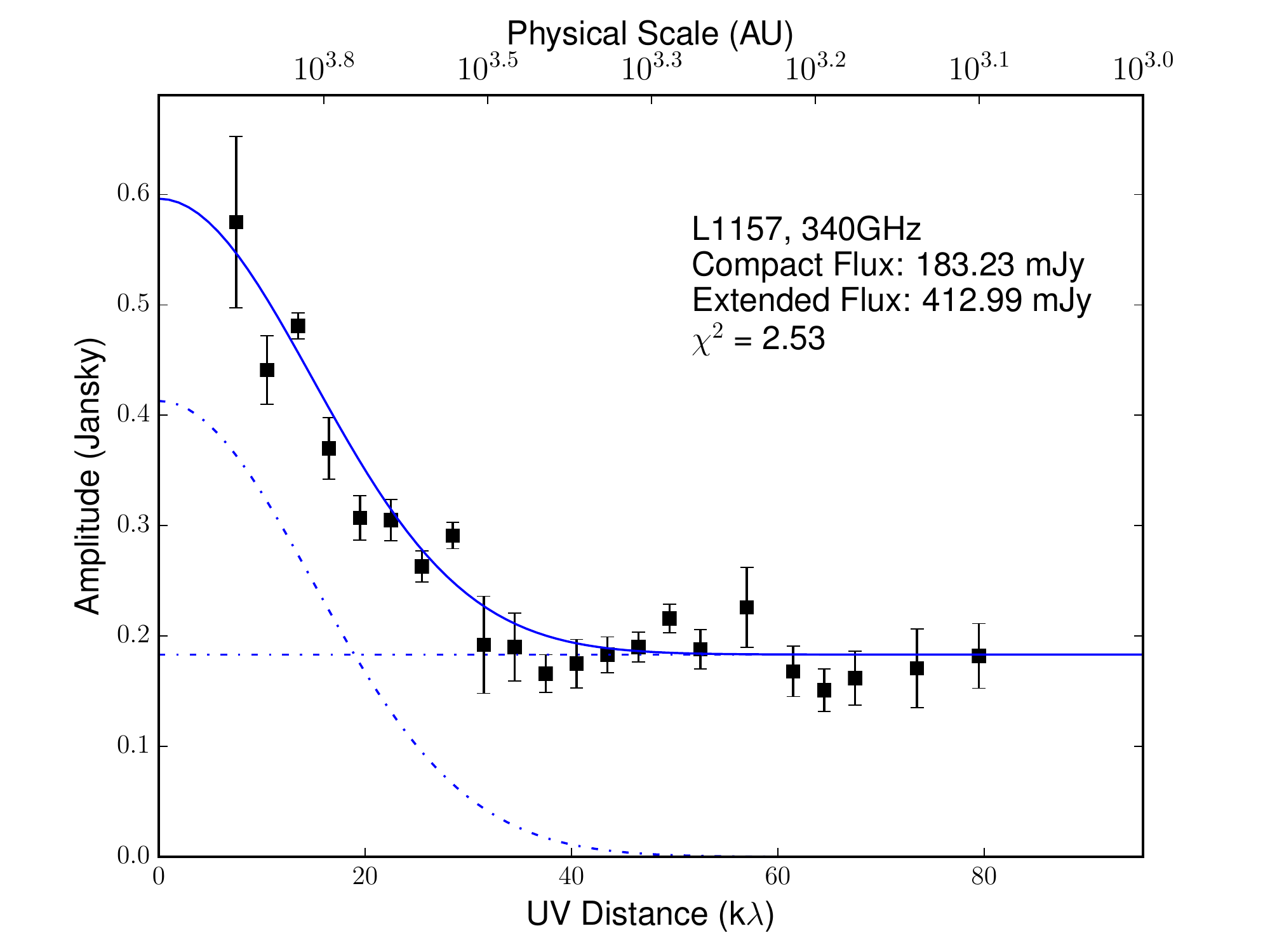}    
\caption{An example of the Gaussian decomposition for the data taken toward L1157.
The black squares represent the observed visibility after performing the optimized binning as described in section 3.1. 
While the error bars here are obtained by error propagtion in the binning process, the original uncertainty level is limited by the {\tt uv} sampling of the original observation.
The best fit is shown by the blue solid line, with each components denoted by the blue dot-dashed lines. 
The resultant fluxes for the compact and extended components are described in the upper right corner of the plot.
}
    \label{fig:example}
    \vspace{0.5cm}
\end{figure}

In the actual decomposition process, we adopt two kinds of fitting profiles: the binned visibility amplitudes are fitted by two Gaussian profiles if the compact component is marginally resolved, while they are fitted by the combination of one Gaussian profile and a point source (i.e. a straight line with a constant amplitude) if the compact component is entirely unresolved.
The best fit was searched by evaluating the reduced $\chi^{2}$.
As an example, Figure \ref{fig:example} shows the results for L1157. 
The filled squares with error bars denote the binned visibility amplitudes, and the solid curve is for the best fit. 
For this target, the compact component is fitted well by the straight line (i.e., a point source),
while the extended component is by a Gaussian profile.

\begin{table*}
\caption{Gaussian decomposition in the $uv$-amp domain}
\centering
\begin{tabular}{cccccc}
\toprule
\multirow{2}{*}{Source}	& \multirow{2}{*}{Frequency}	& \multicolumn{2}{c}{Extended Component} & \multicolumn{2}{c}{Compact Component} \\
 & & Flux (mJy)   & Uncertainty (mJy)   & Flux (mJy)    & Uncertainty (mJy)    \\
\midrule
\multirow{4}{*}{L1448CN}		
& 	37  & 0.22     & 0.06     & 2.01      & 0.05    \\[2pt]
&	43  & 0.73   & 0.39  & 1.95   & 0.02 \\[2pt]
&	230 & 344.46 & 15.16 & 136.43 & 3.93 \\[2pt]
&	340 & 275.98 & 49.28 & 305.01 & 6.1  \\[2pt]
\hline
\multirow{5}{*}{NGC1333 IRAS2A}
 & 	37  & --     & --     & 1.55    & 0.08  \\[2pt]
 & 	43  & 1.18     & 0.47     & 1.17   & 0.08  \\[2pt]
 & 	100 & 29.16    & 2.92     & 33.04  & 3.04  \\[2pt]
 & 	194 & 12808.92 & 28263.68 & 323.17 & 32.06 \\[2pt]
 & 	340 & 826.11   & 45.34    & 554.21 & 34.68 \\[2pt]
\hline
\multirow{7}{*}{NGC1333 IRAS4A1}	
 & 37  & 5.75     & 0.65     & 7.83      & 0.29    \\[2pt]
 & 41  & 7.86    & 0.51   & 8.83   & 0.3   \\[2pt]
 & 43  & 6.4     & 0.95   & 11.26  & 0.54  \\[2pt]
 & 45  & 9.85    & 0.89   & 11.81  & 0.81  \\[2pt]
 & 47  & 10.14   & 1.29   & 13.61  & 1.25  \\[2pt]
 & 230 & 2535.59 & 114.3  & 588.62 & 32.62 \\[2pt]
 & 340 & 2177.26 & 216.03 & 897.91 & 46.18 \\[2pt]
\hline
\multirow{3}{*}{NGC1333 IRAS4A2}	
 & 37  & 5.99     & 3.01     & 0.74     & 0.05    \\[2pt]
& 230 & 2821.82 & 165.04 & 225.48 & 31.57 \\[2pt]
& 340 & 2735.19 & 557.3  & 539.36 & 47.32\\[2pt]
\hline
\multirow{4}{*}{Barnard 1b-N}
 & 37  & 1.21     & 0.06     & 0.3      & 0.03    \\[2pt]
 & 230 & 67.17    & 24.88    & 80.61    & 7.63    \\[2pt]
 & 280 & 196.26   & 42.56    & 232.16   & 20.17   \\[2pt]
 & 340 & 667.1    & 445.19   & 415.05   & 27.3   \\[2pt]
\hline
\multirow{5}{*}{Barnard 1b-S}
 & 37  & 0.67    & 0.05   & 0.24   & 0.03  \\[2pt]
 & 100 & 19.04   & 7.57   & 18.17  & 2.54  \\[2pt]
 & 230 & 68.03   & 38.22  & 249.89 & 7.03  \\[2pt]
 & 280 & 1174.25 & 397.04 & 378.92 & 8.15  \\[2pt]
 & 340 & 337.44  & 101.03 & 811.71 & 30.71\\[2pt]

\hline
\multirow{4}{*}{L1527}  
 & 43  & 2.33   & 0.8   & 1.14   & 0.11  \\[2pt]
 & 90  & 14.37  & 3.84  & 13.45  & 4.05  \\[2pt]
 & 230 & 170.65 & 6.47  & 27.05  & 4.75  \\[2pt]
 & 340 & 283.00 & 24.58 & 267.05 & 26.3 \\[2pt]
\hline
\multirow{3}{*}{Serpens FIRS1}  
 & 43  & 7.33    & 1.48   & 1.5     & 0.23  \\[2pt]
 & 230 & 732.26  & 123.28 & 319.47  & 74.44 \\[2pt]
 & 340 & 3057.06 & 91.43  & 1869.91 & 33.67 \\[2pt]
\hline
\multirow{3}{*}{L1157} 
 & 43 & 6.11 & 0.99 & 0.96 & 0.17 \\[2pt]
 & 230 & 90.68  & 42    & 134.82 & 13.02 \\[2pt]
 & 340 & 412.99 & 37.42 & 183.23 & 9.81  \\[2pt]
\bottomrule
\end{tabular}
\label{tab:uvampfit}
\vspace{0.5cm}
\end{table*}

We found that the FWHM values derived from Gaussian fitting are highly uncertain and cannot be used for quantitative analysis.
The large uncertainties in the size of the compact component originate from the lack of long baseline data that are necessary to resolve the source.
Nonetheless, the obtained FWHM is useful for confirming that the emission of the compact component originates from the roughly same physical scale at all the frequency band observations. 

\begin{figure*}[t]
\centering
	\scalebox{0.80}{
	\begin{tabular}{@{}cc@{}}
	\includegraphics[width=0.4\textwidth]{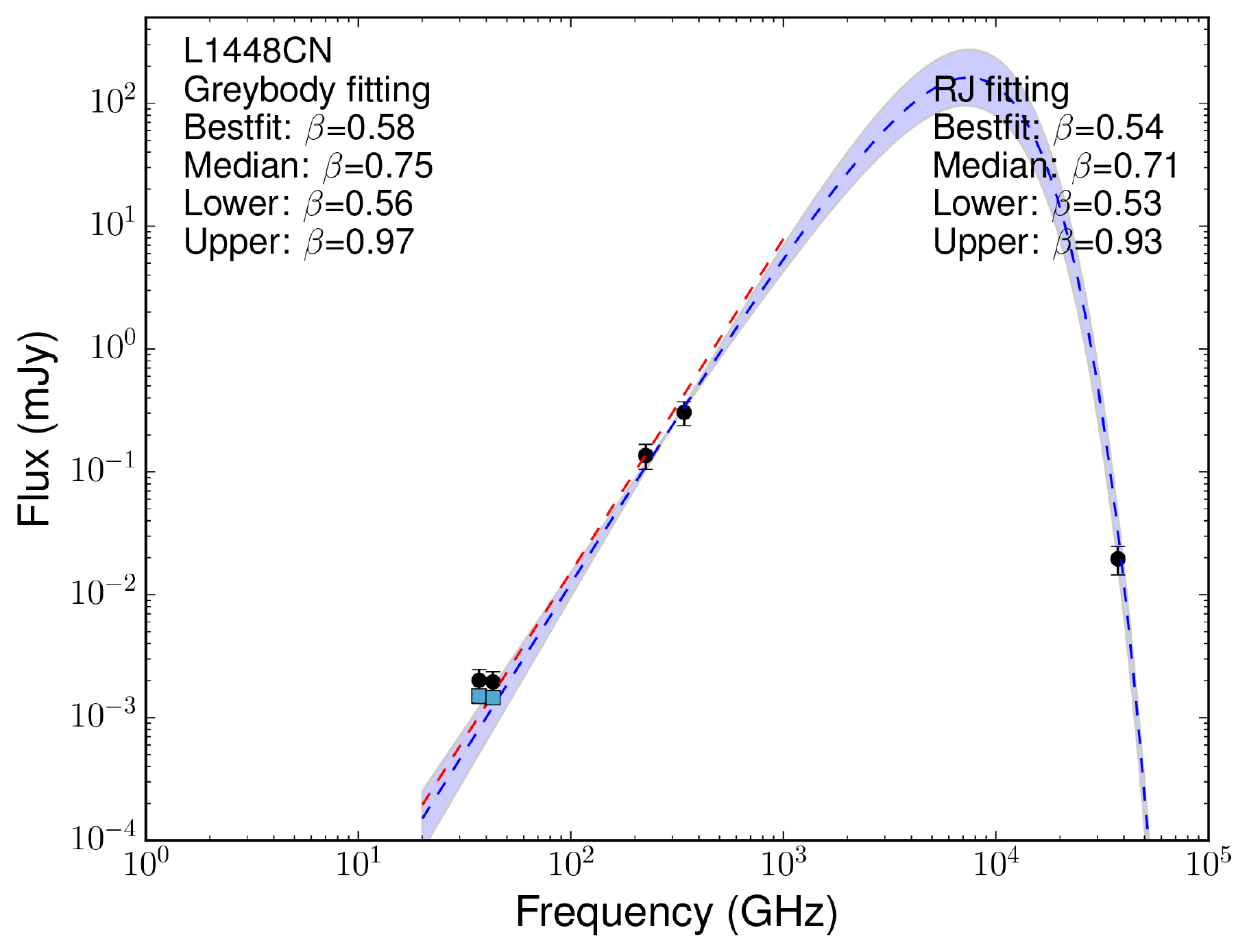}
        \includegraphics[width=0.4\textwidth]{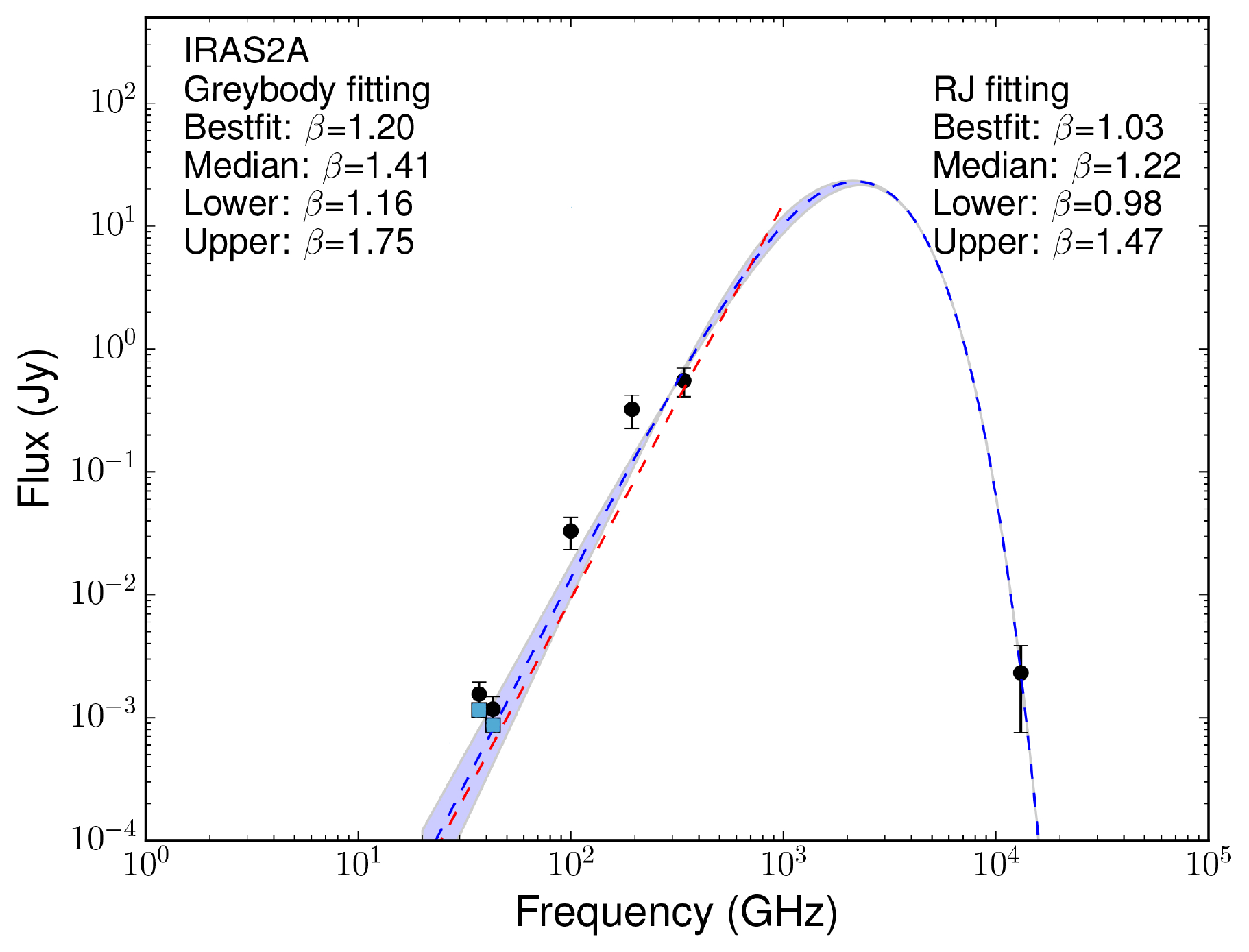}
        \includegraphics[width=0.4\textwidth]{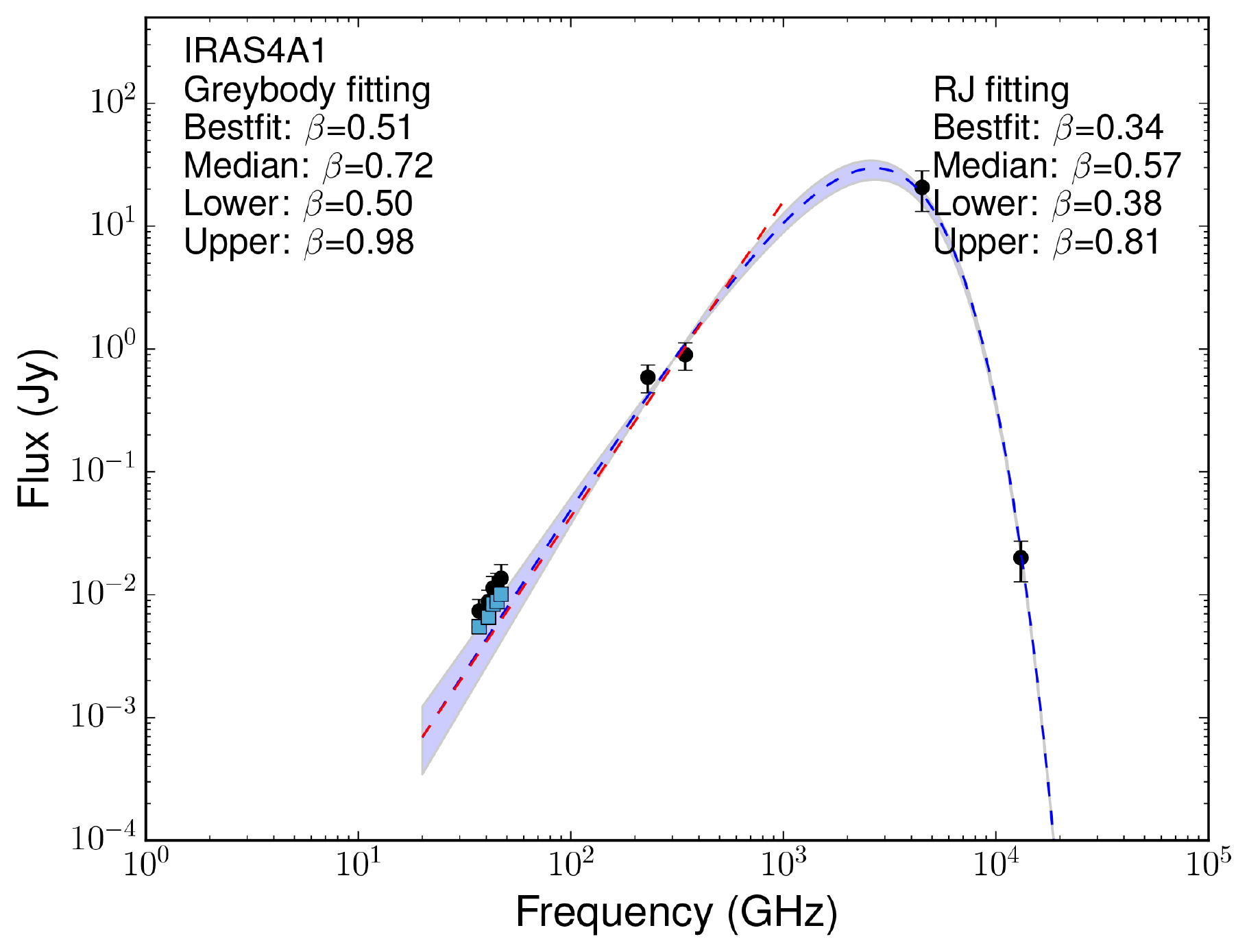}\\
        \includegraphics[width=0.4\textwidth]{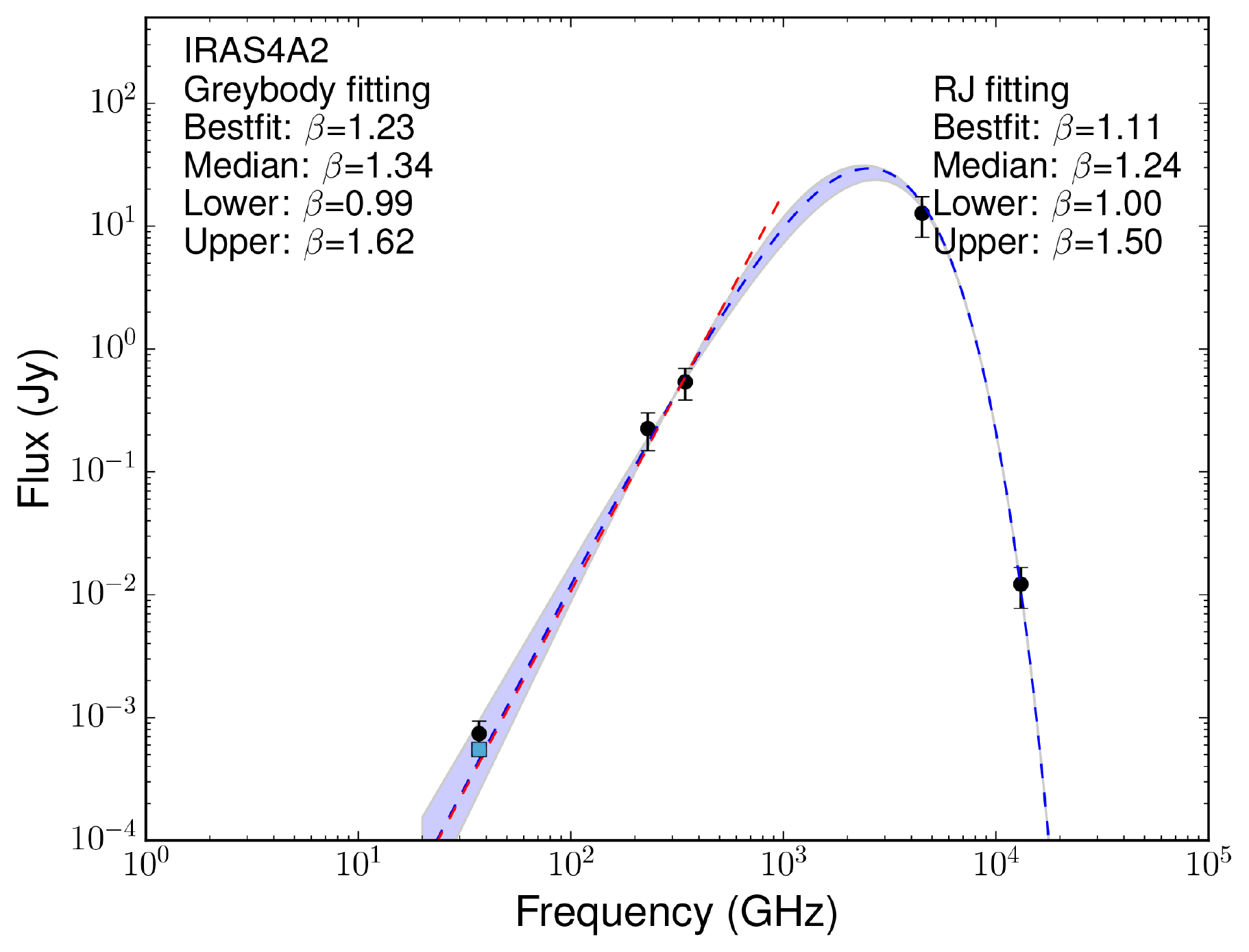}
	\includegraphics[width=0.4\textwidth]{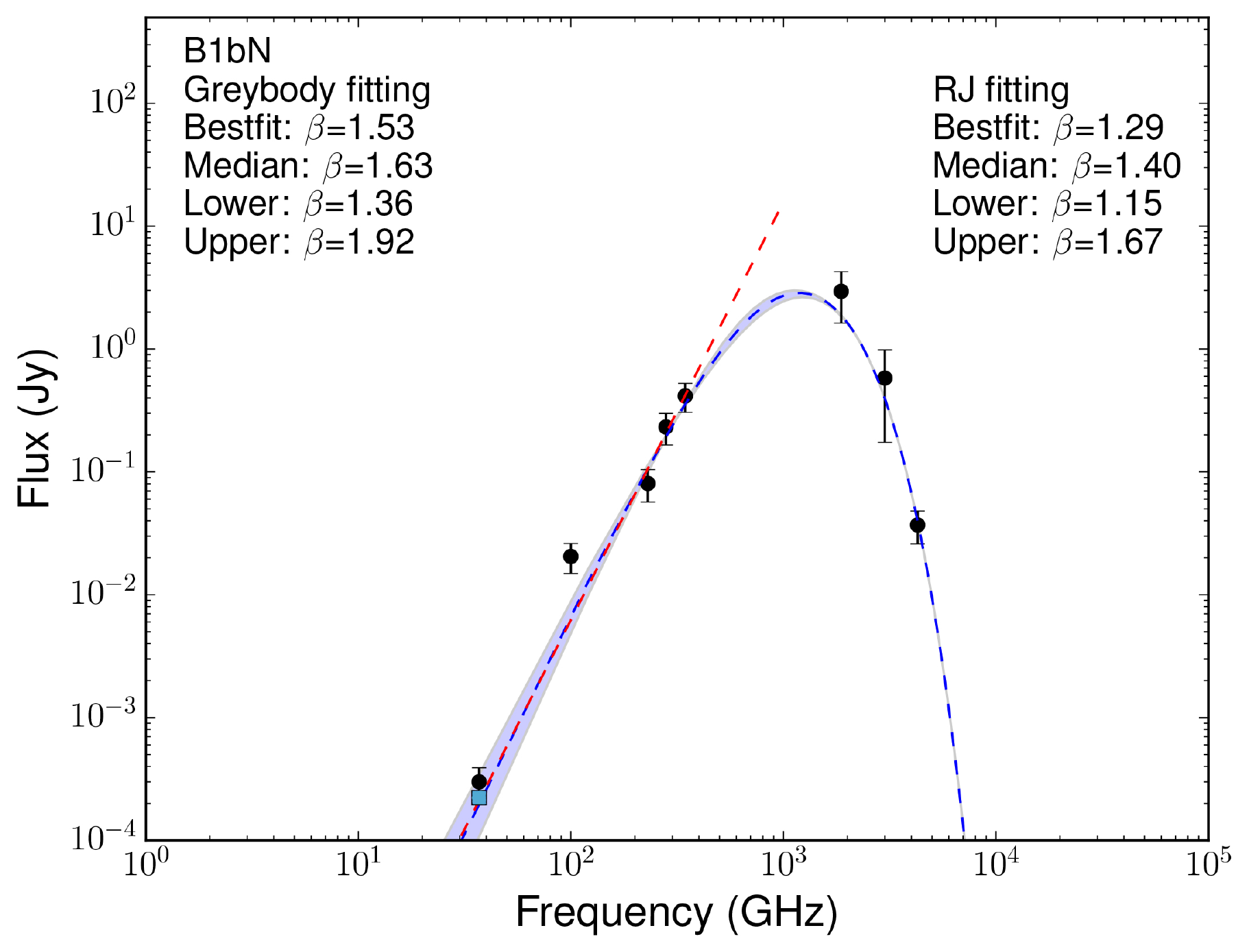}
        \includegraphics[width=0.4\textwidth]{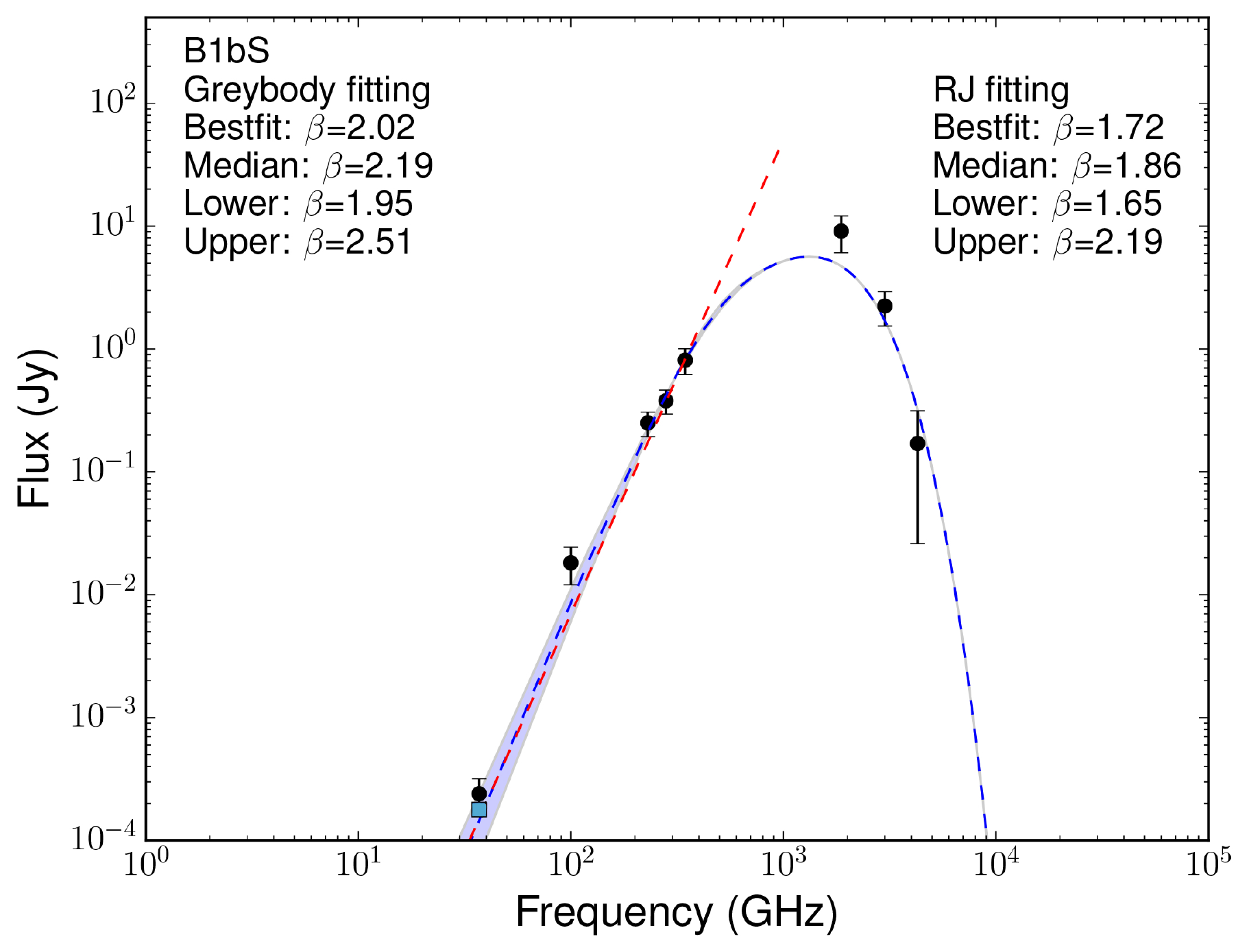} \\
        \includegraphics[width=0.4\textwidth]{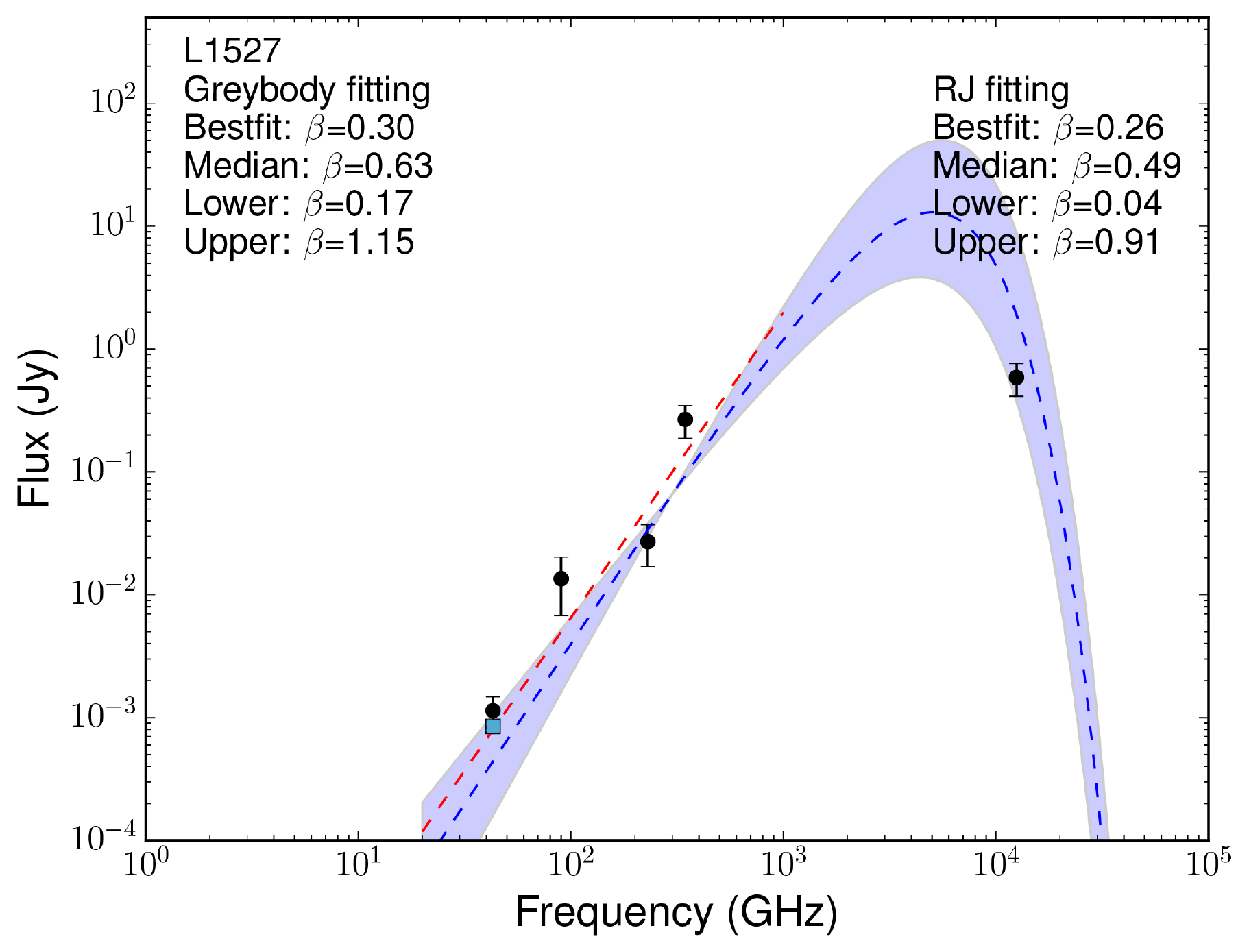} 
        \includegraphics[width=0.4\textwidth]{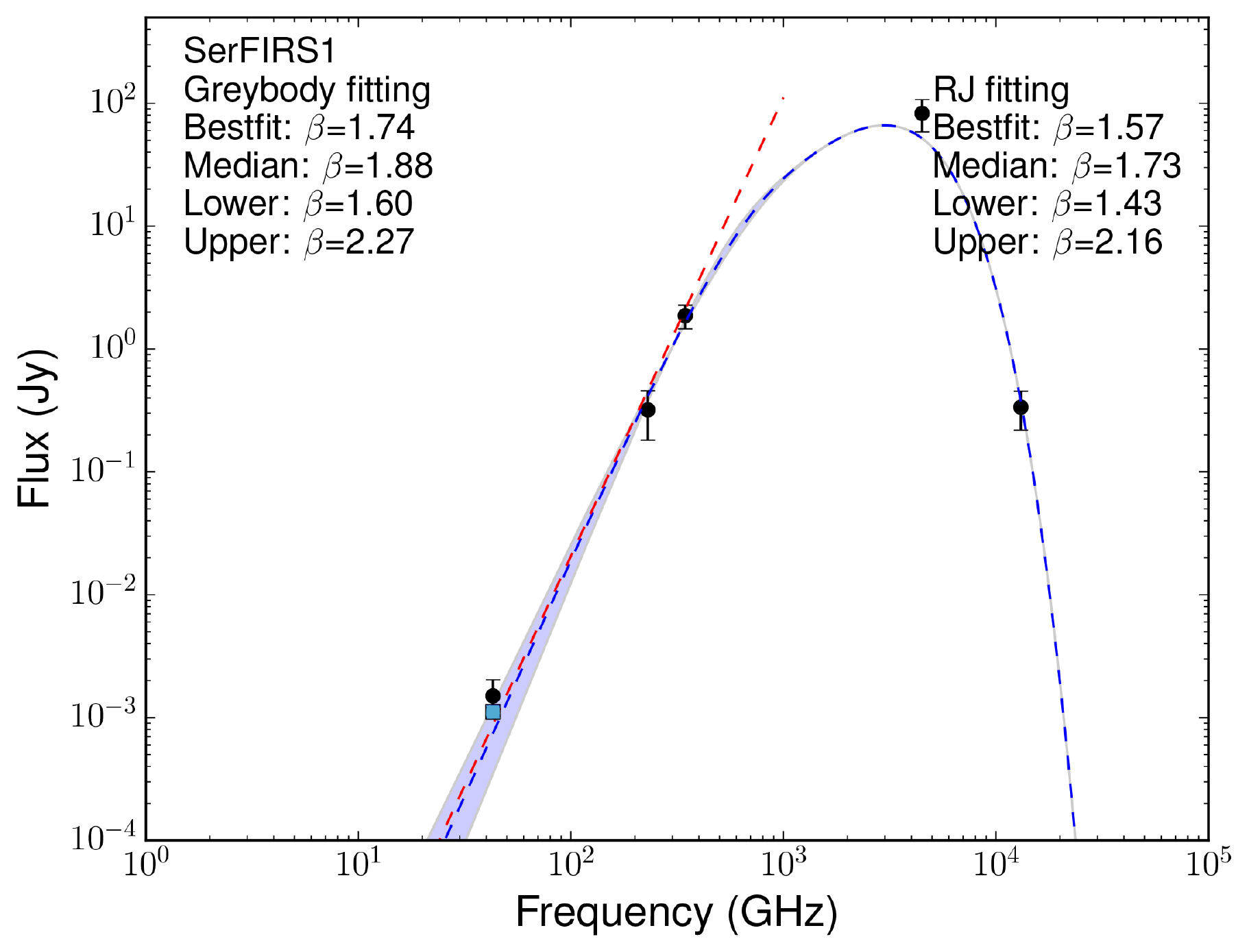}
        \includegraphics[width=0.4\textwidth]{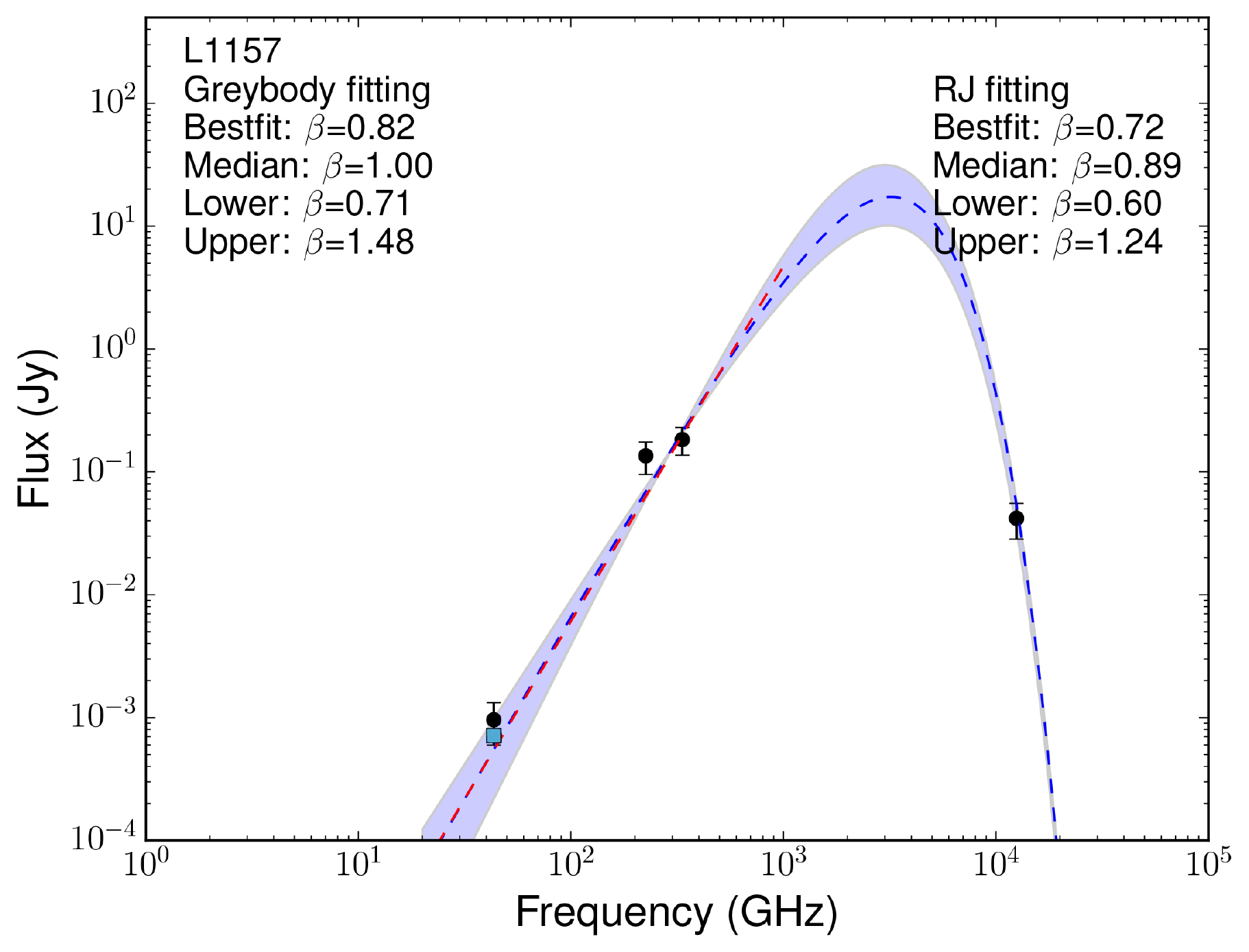}\\
	\end{tabular}}
    \caption{The resultant SEDs for our target sources. 
    Black data points are the fluxes of the compact components derived from our Gaussian decomposition and the fluxes measured by the Spitzer MIPS observations. 
{Blue} squares denote the corrected fluxes assuming that 25$\%$ of the total flux arises from free-free emission.
In other words, {almost all fluxes measured in the millimeter and submillimeter} can be regarded as the ones originating purely from the dust thermal emission. 
{The} red dashed line represents the median value of $\beta_{RJ}$ obtained from SED fitting with the optically thin, Rayleigh-Jeans approximation, {and} the blue dashed curve is from the full modified black body, single emission component SED fitting. 
Note that the MC error estimation is carried out for both the cases as described in Section \ref{sub:freefree}. 
The blue shaded areas show the {SEDs with} upper and lower limits of $\beta$.}
    \label{fig:sed}
    \vspace{0.5cm}
\end{figure*}

\subsection{SED fitting}\label{sub:sed}
Our assessment for the results of Gaussian decomposition is that the extended components do not have a well defined size and geometry, and are seriously subject to different degrees of missing fluxes at different frequency bands.
Therefore, the determination of the total flux for the extended components is highly sensitive to the shortest {\it uv} spacing in the interferometric observations.
Since we do not have short spacing data from single dish observations for most of our targets, we only make use of the compact components for all the targets in the following SED fitting.

The compact components in general have much higher brightness than the extended components, and their observed fluxes are not sensitive to the missing flux problem.
Furthermore, the compact components correspond to much higher density regions of YSOs and thereby a higher degree of grain growth can be expected there, compared with the extended components.
Thus, analyzing the compact components is more relevant to our purpose of studying the initial dust grain growth in star-forming regions (e.g., \citealt{Wong2016}).
Therefore, we perform modified black body fittings to the fluxes of the compact components to derive the dust temperature, optical depth and $\beta$ simultaneously. 

To proceed, we assume that the flux ($F_{\nu}$) observed at a frequency ($\nu$) can be expressed by the modified black body emission \citep{hildebrand1983}:
\begin{equation}
\label{eq:F_nu}
F_{\nu} =\Omega(1-{e}^{-\tau_{\nu}})B_{\nu}(T),
\end{equation}
where $\Omega$ is the solid angle related to the physical size of a target source ($r=\sqrt{ln2\times\Omega/\pi}$), $\tau_{\nu}$ is the optical depth measured at $\nu$ from the observers towards the source, and $B_{\nu}(T)$ is the Planck function for a certain temperature ($T$) of the source.
We assume that compact components at all frequency bands share an identical sold angle of $\Omega$=0$\farcs$5, which can be factored out during the derivation of beta. 
A simple power-law profile is adopted for $\tau_{\nu}$:
\begin{equation}
\label{eq:taudust}
 \tau_{\nu} = \tau_{0} \left( \frac{\nu}{\nu_0}\right)^{\beta} 
 \left( \mbox{or } \kappa_{\nu} = \kappa_0 \left( \frac{\nu}{\nu_0}\right)^{\beta}  \right),
\end{equation}
where $\tau_0$ is the optical depth at $\nu=\nu_0$. We set $\nu_0=$ 300 GHz in our fittings. 
In the case of spatially uniform $\kappa_{\nu}$, $\tau_{\nu}$=$\kappa_{\nu}$$\int\rho d\ell$=$\kappa_{\nu}$$\Sigma$, where $\Sigma$ is dust mass surface density, and $\ell$ is the thickness of structure along the line-of-sight.

{In the case where} dust emission is optically thin (i.e. $\tau_{\nu} \ll 1$), equation (\ref{eq:F_nu}) can be simplified according to ${e}^{-\tau_{\nu}} \simeq 1 - \tau_{\nu}$.
In addition, when the observed frequency bands are in the Rayleigh-Jeans limit ($h \nu/KT \ll 1$), then $B_{\nu}(T)$ becomes approximately proportional to $\nu^2$.
Consequently, $F_{\nu}$ can be simplified as 
\begin{equation}
\label{eq:F_nu_rj}
F_{\nu} \propto \nu^{2+\beta_{RJ}}
\end{equation}
We label $\beta$ as $\beta_{RJ}$ to emphasize that the value of $\beta_{RJ}$ is derived under the assumption of the optically thin emission and the Rayleigh-Jeans limit. 
Based on these assumptions, equation (\ref{eq:F_nu_rj}) permits deriving $\beta_{RJ}$ from observations at merely two frequency bands.
{The} approach of $\beta_{RJ}$ has been widely adopted in the literature for examining grain growth.
{ However, when using only two frequency bands (e.g. $\sim$230GHz and $\sim$340GHz), the small separation in the frequency range and uncertainties from flux measurements would yield a large uncertainty in the derivation of beta. For instance, considering same uncertainty estimation as described in Section \ref{sub:freefree}, $\beta$ derived from $\sim$230GHz and $\sim$340GHz bands give uncertainties of $>$1 for all sources in our sample, and can not be used as a tool to diagnose grain growth as in literatures.}

{Also}, it should be noted that many Class 0 objects tend to be highly obscured even at (sub)millimeter wavelengths and can have a relatively low average dust temperature (e.g., $<$ 20 K). 
In these cases, both the optically thin and the Rayleigh-Jeans assumptions break down.
This may be particularly significant for the observations at $\sim$340 GHz or higher frequencies.
In this work, we hence quantify the difference between the values of $\beta$ and $\beta_{RJ}$ to compare them with each other.

\subsection{Uncertainty estimation}\label{sub:freefree}

To realistically determine the uncertainty of $\beta$ and $\beta_{RJ}$, we perform a Monte Carlo (MC) error estimation. 
In this procedure, we first estimate the uncertainty of the observed fluxes that is caused by errors arising from a number of sources. 
In this paper, we consider two plausible sources: the intrinsic error coming from observations and data analyses, and the confusion between the dust and the non-dust emissions. 
Once the flux uncertainty is determined for these two sources (see below), we then {alter} the measured flux randomly based on the flux uncertainty estimation. 
We {use} the perturbed fluxes to perform SED fitting and obtain the resultant value of $\beta$ and $\beta_{RJ}$. 
The above procedure is repeated for 1000 times to compute the median values of $\beta$ and $\beta_{RJ}$. 
We define the median value as the most likely value of $\beta$ and $\beta_{RJ}$ to explore a possibility of dust grain growth in Class 0 YSOs (see Section \ref{sec:result}). 
{The median is preferred over the mean because the distributions from MC error estimation are not always symmetric and the median is less affected by the MC realization with extremely steep or flat SEDs. Generally, the difference between the mean and median values are within the uncertainty given for $\beta$ and $\beta_{RJ}$.}
For the uncertainty, we adopt the 16 and 84 percentile values from the 1000 repetitions as the lower and upper limits.
Furthermore, we conduct the reduced $\chi^2$ analysis to find out the best fit for $\beta$ under that assumption that free-free emission contributes to 25\% of the observed flux at longer wavelengths (see Table \ref{tab:results}).

\subsubsection{Intrinsic errors}
We adopt the flux measurement uncertainty from the Gaussian decomposition procedure ($\sim$10\%) as the statistical flux measurement error (not including absolute flux uncertainty).
In addition, we consider the nominal, 20\% absolute flux calibration uncertainty for all of our observations by adding it onto the statistical flux error.
{For each MC repetition, we incorporate the intrinsic errors by drawing the perturbed fluxes from Gaussian distributions with the measured flux as the mean and the uncertainties as the standard deviation.}

\subsubsection{Contribution from non-dust emission}\label{subsub:freefree}
For Class 0/I objects, free-free emission may be produced by the $\sim$100 AU scale thermal radio jets {\citep[e.g.,][]{Rodriguez1997}}, and occasionally {by} the spatially unresolved non-thermal flares \citep{Forbrich2006} {for} which spectral indices are not yet certain \citep{Liu2014}.
The confusion between the dust and the non-dust emissions due to these phenomena may be relatively important at 43 GHz or lower observing frequencies especially for the compact component.
Without considering such confusion, SED fitting may lead to underestimating $\beta$.
The SED measured at centimeter wavelengths can constrain such confusion.
However, this is not always possible when utilizing the archival data which were taken at different times, since the fluxes of radio jets and the non-thermal flares both vary with time {\citep[for the characteristic timescale, see][]{Liu2014}}.

For a Class 0/I object at $d \sim 140$ pc, the typical radio jet flux may be on the order of 0.1--1 mJy, and the typical non-thermal emission (if present) may be $\sim$0.1 mJy or fainter \citep[e.g.,][]{Dzib2013,Liu2014,Pech2016}.
Our selected target sources appear to be embedded in relatively {abundant} circumstellar gas and dust, and present brighter 37 GHz and 43 GHz emission than the expected fluxes of optically thin non-dust emission (see Table \ref{tab:uvampfit}).
For the case of L1448 CN, \citet{Hirano2010} found that free-free emission can contribute {to} up to $\sim$50\% of {the} flux at 30--40 GHz.
Based on these observations, we empirically assume in our SED fitting that the non-dust emission may contribute to 0--50$\%$ of the total 37 GHz or 43 GHz fluxes at the epoch of those observations. 
For each MC repetition, we randomly assign a non-dust emission factor of 0.0--0.5 to the 30--43 GHz fluxes and subtract the computed non-dust emission to evaluate a potential bias in our determination of $\beta$. 
{Since the properties of non-thermal emission in many of our sources are uncertain, the non-dust emission factor is drawn from a uniform distribution.}
Non-dust emission is negligible at higher frequencies.  
This correction only lowers the fluxes by a factor of two at most.
As a result, (potential) contributions from the non-dust emission yield negligible effects in the derived $\beta$ thanks to the wide frequency and flux range of the SEDs.

\section{Results}\label{sec:result}
\begin{table*}
\caption{The results of our SED fitting}  
\label{tab:results}
\centering
\scalebox{0.9}{
\begin{tabular}{l|c>{\bf}ccc|cccc|cccc|c>{\bf}ccc}
\hline\hline
&&&&&&&&&&&&&&&&\\[0.1pt]
Source  & \multicolumn{4}{c|}{$\beta$} & \multicolumn{4}{c|}{Temperature} & \multicolumn{4}{c|}{$\tau_{1mm}$} & \multicolumn{4}{c}{$\beta_{RJ}$} \\[2pt]
& best fit &median & $16\%$ & $84\%$ & best fit &median & $16\%$ & $84\%$  & best fit &median & $16\%$ & $84\%$& best fit &median & $16\%$ & $84\%$ \\[2pt]

\hline
&&&&&&&&&&&&&&&&\\[0.1pt]
L1448CN  & 0.58 & 0.75 & 0.56 & 0.94 & 102 & 99 & 97 & 103 & 0.03 & 0.03 & 0.03 & 0.04 & 0.54 & 0.71 & 0.52 & 0.90 \\[1pt]
IRAS2A   & 1.20 & 1.41 & 1.16 & 1.75 & 36  & 36 & 34 & 37  & 0.24 & 0.24 & 0.17 & 0.32 & 1.03 & 1.22 & 0.98 & 1.47 \\[1pt]
IRAS4A1  & 0.51 & 0.72 & 0.50 & 0.98 & 41  & 41 & 40 & 42  & 0.37 & 0.33 & 0.24 & 0.44 & 0.34 & 0.57 & 0.38 & 0.81 \\[1pt]
IRAS4A2  & 1.23 & 1.34 & 0.99 & 1.62 & 40  & 40 & 38 & 40  & 0.18 & 0.16 & 0.09 & 0.21 & 1.11 & 1.24 & 1.00 & 1.49 \\[1pt]
B1bN     & 1.53 & 1.63 & 1.36 & 1.92 & 18  & 18 & 17 & 19  & 0.32 & 0.30 & 0.21 & 0.39 & 1.29 & 1.40 & 1.15 & 1.67 \\[1pt]
B1bS     & 2.02 & 2.19 & 1.95 & 2.51 & 22  & 22 & 19 & 23  & 0.61 & 0.60 & 0.45 & 0.84 & 1.72 & 1.86 & 1.65 & 2.18 \\[1pt]
L1527    & 0.30 & 0.63 & 0.17 & 1.15 & 69  & 58 & 51 & 76  & 0.01 & 0.02 & 0.01 & 0.03 & 0.26 & 0.49 & 0.04 & 0.91 \\[1pt]
SerFIRS1 & 1.74 & 1.88 & 1.60 & 2.27 & 51  & 51 & 49 & 52  & 0.39 & 0.36 & 0.25 & 0.48 & 1.57 & 1.73 & 1.43 & 2.16 \\[1pt]
L1157    & 0.82 & 1.00 & 0.71 & 1.48 & 43  & 42 & 41 & 44  & 0.06 & 0.06 & 0.04 & 0.08 & 0.72 & 0.89 & 0.60 & 1.23 \\[1pt]
\hline
\end{tabular}}
\vspace{0.5cm}
\end{table*}

\begin{figure}
\hspace{-0.2cm}
\includegraphics[width=8.8cm]{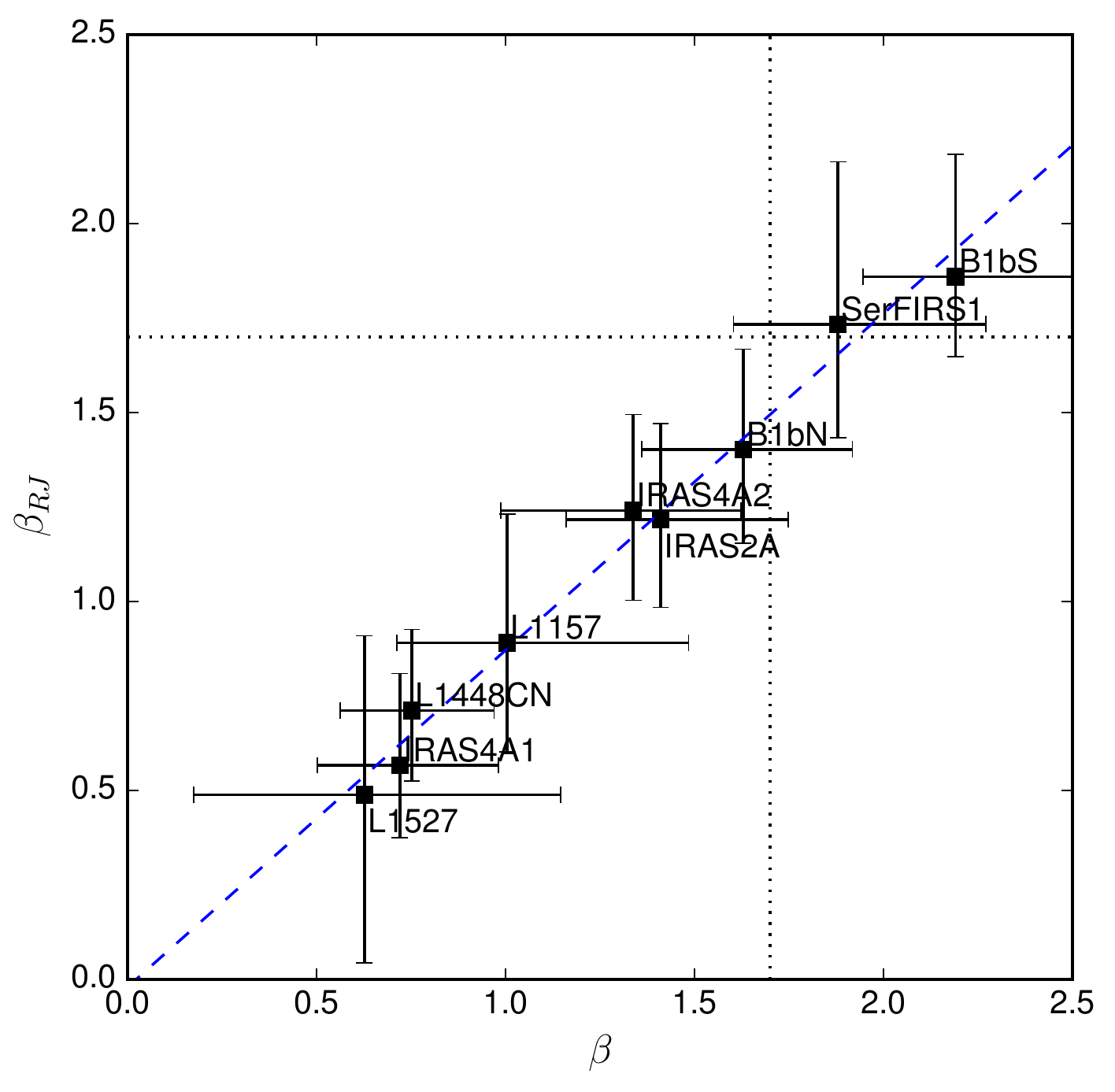}
\caption{
Distribution of $\beta$ and $\beta_{RJ}$ for all the targets (see the filled squares with error bars).
The black dotted lines denote the typical value of beta for the interstellar dust ($\beta \sim 1.7$). The blue dashed line represents the best fit for a correlation between $\beta$ and $\beta_{RJ}$.
}
\label{fig:2beta}
\end{figure}

\subsection{{Dust opacity indices $\beta$}}

Table \ref{tab:uvampfit} summarizes the fluxes of the spatially compact components that are derived from our Gaussian decomposition (Section \ref{sec:data_ana_2}).
The results of SED fitting for the compact components are plotted in Figure \ref{fig:sed}. 
The obtained values of $\beta$, $\beta_{RJ}$, the dust temperature, and the optical depth are listed in Table \ref{tab:results}.
A comparison between the values of $\beta$ and $\beta_{RJ}$ is shown in Figure \ref{fig:2beta}.

Our results show that both $\beta$ and $\beta_{RJ}$ distribute widely, with the average values of $\beta= 1.3 $ and $\beta_{RJ} = 1.1 $ (Figure \ref{fig:2beta}). 
The derived $\beta$ of SerFIRS\,1, B1-bN, IRAS\,2A and IRAS\,4A2 are comparable to the interstellar value of $\sim$1.7, 
while the $\beta$ of B1-bS is slightly larger than 2.0. 
For the rest of our targets, the resultant values of $\beta$ are $<$1.0. 
In addition, Figure \ref{fig:2beta} shows that the derived $\beta_{RJ}$ and $\beta$ tend to be well correlated with each other. 
A linear regression gives 
\begin{equation}
\beta_{RJ} = (0.89\pm0.002) \times\beta-(0.03\pm0.004).
\end{equation}
We also find that the fitted optical depths at $\lambda$$=$1 mm range from 0.02 to 0.61 and the resultant dust temperatures are in the range of from $\sim$10 K to $\sim$100 K for our targets (see Table \ref{tab:results}).

The fitted values of $\beta$ for our samples are broadly in agreement with the previous studies. 
For instance, our results indicate that B1-bN and B1-bS have larger values of $\beta$ and the lowest dust temperatures in our targets, 
which are both consistent with the previous estimate done by \cite{hirano2014}. 
As other examples, the previous derivations based on CARMA observations have suggested that L1157 has $\beta$$\sim$0.5 \citep{Kwon2009, chiang2012envelope}, and that L1527 has $\beta$$\sim$0 \citep{tobin2013modeling}.  
\cite{Jorgensen2009} also investigated the dust opacity indices for L1448C, IRAS\,2A, IRAS\,4A, IRAS\,2B, L1157 and L1527 using the SMA data at 0.88 and 1.3 mm, 
and found that the calculated $\beta$ are around 0--1 at long baselines ($>$ 40k$\lambda$) for all their sources. 
Such low values of $\beta$ are consistent with our single-component fittings. 

We caution that the derived, lower than 1.7 values of $\beta$ here do not immediately imply the evidence of grain growth.
The values of $\beta$ can be underestimated when the assumption of a single uniform emission component breaks down.
In fact, a signature of this possibility can be seen in Figure \ref{fig:sed};  
the observed $\sim$340 GHz fluxes of L1448CN, IRAS\,2A, IRAS\,4A1, and L1157 are smaller than the regression lines derived from the optically thin and Rayleigh-Jeans assumptions. 
If such differences are not entirely due to measurement errors, then these differences suggest that the dust emission at $\gtrsim$230 GHz should already become optically thick for these targets. 
This is however inconsistent with the low $\tau_{\mbox{\scriptsize{1\,mm}}}$ values given by the full single-component modified black body fittings, especially for the cases of L1448CN and L1157 (Table \ref{tab:results}).
In addition, we have to admit that the observed SED of L1527 is not fitted well by either the optically thin, Rayleigh-Jeans approximation or the full modified black body fitting with the single emission component (see Figure \ref{fig:sed}).
Furthermore, while some of the previous studies concluded that the small $\beta$ value indicates (a significant degree of) grain growth, \cite{Jorgensen2009} also pointed out that resolved observations of the compact region and modeling of the envelope and disk structure are necessary to put tighter constrains on $\beta$ ({see Section \ref{sec:tempandbeta} for more quantitative discussion)}. 
Finally, the recent, higher angular resolution VLA observations show that the $\beta$ of IRAS 4A1 and 4A2 are much closer to $\sim$1.7 and $\sim$1.2 between 8 and 10 millimeter wavelengths, respectively \citep{cox2015VANDAM}. It is obvious that there is a big difference between their estimate and the one done by us and \citet{Jorgensen2009} for IRAS 4A1.
We will discuss below a potential origin of why such inconsistencies emerge {(see Section \ref{sec:toymodel}).}

\subsection{{Degeneracy between dust temperature and $\beta$}}\label{sec:tempandbeta}

FIR data points are included in our SED fitting to provide a better constraint on the estimate of the dust temperature ($T$). 
However, due to the low spatial resolution of available observations at these wavelengths, we cannot distinguish the compact and extended components for the observations, as done for our radio/submillimeter data. 
The mixture of the flux from both the components can result in underestimating $T$ and $\beta$ for the compact component. 
We quantify this effect on $\beta$ by fitting SEDs at centimeter to submilimeter wavelengths with $T$ fixed at 100K. Note that the value of $T=100$ K is a conservative upper limit since the averaged dust temperature around these sources is generally $<$60K. 
We find that for this case $\beta$ is underestimated only by $\sim$10$\%$ compared to $\beta_{RJ}$, which is on the same scale of the our error estimation. 
Thus, we can conclude that the low spatial resolution observations at FIR do not affect our estimate of $\beta$ very much.

\subsection{{Effects of multiple emission components to the $\beta$ measurements}}\label{sec:toymodel}

\begin{table*}
\caption{Parameters for three-components SED models}  
\centering
\label{tab:3components}
\begin{tabular}{ | l | c c c | c c c| c c c | }\hline\hline
  & & & & & & & & &   \\[0.2pt]
	&  $T_{di}$	& $\Omega_{di}$ & $\kappa_{\mbox{\tiny{230 GHz}}}$$\Sigma^{\mbox{\tiny{di}}}$ & $T_{\mbox{\tiny{do}}}$ & $\Omega_{\mbox{\tiny{do}}}$ & $\kappa_{\mbox{\tiny{230 GHz}}}$$\Sigma^{\mbox{\tiny{do}}}$ & $T_{\mbox{\tiny{ei}}}$ & $\Omega_{\mbox{\tiny{ei}}}$ & $\kappa_{\mbox{\tiny{230 GHz}}}$$\Sigma^{\mbox{\tiny{ei}}}$ \\[5pt]
    & (1) & (2) & (3) & (4) & (5) & (6) & (7) & (8) & (9) \\[5pt]
    & (K) & (sr) &  & (K) & (sr) &  & (K) & (sr) & \\[5pt]\hline
  & & & & & & & & &   \\[1pt]
L1527   & 650 & 1.0$\times$$10^{-13}$ & 9.0 & 12 & 8.0$\times$$10^{-13}$ & 2.8 & 35 & 8.6$\times$$10^{-11}$ & 6.4$\times$$10^{-3}$ \\[6pt]
NGC\,1333\,IRAS4A1 & $\cdots$ & $\cdots$ & $\cdots$ & 45 & 1.2$\times$$10^{-11}$ & 13 & 20 & 2.0$\times$$10^{-11}$ & 1.5  \\[6pt]
L1157   & $\cdots$ & $\cdots$ & $\cdots$ & 35 & 5.8$\times$$10^{-12}$ & 1.7 & 15 & 6.3$\times$$10^{-12}$ & 1.5 \\[6pt]
L1448CN & $\cdots$ & $\cdots$ & $\cdots$ & 40 & 1.4$\times$$10^{-12}$ & 1.5 & 20 & 6.6$\times$$10^{-12}$ & 1.5  \\[6pt] 
NGC\,1333\,IRAS2A & $\cdots$ & $\cdots$ & $\cdots$ & 150 & 5.2$\times$$10^{-12}$ & 0.39 & 15 & 2.0$\times$$10^{-11}$ & 1.0  \\[6pt] \hline
\end{tabular}
\vspace{0.2cm}
\begin{itemize}  %
\footnotesize
\item[] \par We assumed a dust opacity spectral index $\beta$=2.0 for all dust components. We assumed that the contribution from free-free emission is negligible for L1527, NGC1333\,IRAS\,4A1, and L1157.
Model of L1448CN was incorporated with a free-free emission component, with $T_{e}$$=$8,000 K, $EM$$=$1.3$\times$10$^{9}$ cm$^{-6}$\,pc, $\Omega_{ff}$$=$1.7$\times$10$^{-14}$ sr.
1 sr $\sim$4.25$\times$10$^{10}$ square arcsecond.
We note that the contribution from free-free emission is time variable, which needs to be cautioned when comparing these models with the presented centimeter band data or other future observations.
\par (1) Dust temperature of the inner disk component. (2) Solid angle of the inner disk component. (3) The dust mass surface density multiplied by the dust opacity at 230 GHz for the inner disk component, which is dimensionless.  (4) Dust temperature of the outer disk component. (5) Solid angle of the outer disk component.  (6) The dust mass surface density multiplied by the dust opacity at 230 GHz for the outer disk component. (7) Temperature of the inner envelope component. (8) Solid angle of the inner envelope component. (9) The dust mass surface density multiplied by the dust opacity at 230 GHz for the inner envelope component.
\end{itemize}
\vspace{0.6cm}
\end{table*}





Figure \ref{fig:2beta} shows that grain growth has not yet significantly lowered the values of $\beta$ for five of the nine selected target sources.
{Here, we construct a toy model to discuss} whether or not the small values of the derived $\beta$ for the remaining four sources (L1448CN, IRAS4A1, L1157, L1527) indicate dust grain growth.
We will also attempt to interpret some features in the observed SEDs of IRAS2A, which are difficult to be explained only by the effects of grain growth or optical depth when assuming single emission component.

Many of the observed sources show unexpectedly low spectral slopes ($\alpha$) in certain frequency ranges, which correspond to $\alpha$$<$2.0.
Examples include the observations of L1527 in between 90 GHz and 230 GHz, which is consistent with the data presented in \citet{tobin2013modeling}.
Others are L1157, IRAS2A, IRAS4A1, in between $\sim$200 GHz and 345 GHz (see Table \ref{tab:uvampfit}).
Some low values of $\alpha$ ($<$2.0) were also reported by \citet{Jorgensen2009} and \citet{miotello2014}.

One possibility to explain the low $\alpha$ values may be a very low dust temperature (e.g., $\lesssim$10 K), which shifts SED peaks to submillimeter bands.
For example, \citet{hirano2014} have demonstrated that superimposing the fluxes of an 11 K extended dust component with a 23 K compact dust component may explain the observed SED of Barnard 1-bN on $\sim$2$''$ angular scales.
For our present observations of the compact emission of L1527, IRAS\,4A1, L1157, and L1448CN, it is unlikely that the dominant emission at any band is significantly contributed from a $\lesssim$10 K dust component, given the detected high brightness on small spatial scales.
It is strictly impossible for the case of IRAS\,4A1, since the observed brightness temperature at 44 GHz with $\sim$0$\farcs$5 resolution is already 41 K \citep{Liu2016IRAS4A}.
A high dust temperature may also present in the inner region of IRAS2A, given the $\gtrsim$100 K gas excitation temperature reported in the previous spectral line observations (e.g., \citealt{Coutens2015}).
Therefore, it is less likely that the dominant contribution of the (sub)millimeter band continuum emission of IRAS2A is from $<$10 K dust, in spite of the observed steeply decreasing spectral index in between 230 GHz and 345 GHz frequencies.
From the astrochemistry point of view (e.g., \citealt{Sakai2014}), L1527 appears warm (30--90 K) from the 0$\farcs$5-1$''$ resolution observations, and may require at least three density/temperature components to explain the observed abundance distributions.
The previous Herschel observations of high-$J$ ($J_{u}$$>$13) CO lines towards the Class 0 object BHR\,71 also indicated a very hot ($\sim$400-1000 K) gas component near the embedded protostar {\citep{Yang2017}}.

On the other hand, an obscured, inner hot dust component may only significantly contribute to the observed fluxes at long wavelength bands when it is embedded by an optically thicker, cooler outer component.
This can be a mechanism to produce an excess of strong emission at long wavelength bands, {which is alternative to the common interpretation with significant dust grain growth (i.e., small $\beta$ values), very cold dust, or very optically thick dust.}
In realistic Class 0 YSOs, such a hypothesized geometric picture may be realized as an inner hot (pseudo)disk of few AU scales, obscured by an optically thick outer cool (pseudo)disk to few tens of AU.
For a deeply embedded, high density accretion flow, important heating mechanisms can include compression work, viscous heating, and radiative heating.
However, most of the previous radiative transfer modeling for fitting observed SEDs only consider radiative heating.
An example of temperature and column density profile for embedded YSOs based on numerical simulations can be found in \citet{Vorobyov2013} and reference therein.
Furthermore, Class 0 YSOs may be further embedded within the more extended, optically thinner inner envelope.
Such hypothesized geometric and temperature profiles may be particularly relevant for the case of L1527, since its geometry has been resolved approximately as an edge-on accretion flow \citep{tobin2013modeling}.
The rotating disk was also identified from the high angular resolution VLA observations of NH$_{3}$ lines towards IRAS\,2A \citep{Choi2010}.
{Here we cross reference to a parallel work reporting a spatially resolved case of self-obscured disk at 345 GHz,
which is approximately in an edge-on projection \citep{Lee2017}.}

As motivated by the above consideration, we undertake SED fitting with a toy model of the multiple emission components. In the model, {we suppose that an inner hot pseudo-disk with a scale of a few AU is obscured by an outer, optically thick, cool pseudo-disk with a scale of a few tens of AU}. After incorporating the potential contribution of free-free emission from an ionized disk or a jet ($F^{ff}_{\nu}$), the overall integrated flux of such a system may be approximated as

\begin{equation}\label{eqn:multicomponent}
\begin{split}
F_{\nu} = \Omega_{di}(1-e^{-\tau^{di}_{\nu}})B_{\nu}(T_{di})e^{-\tau^{do}_{\nu}-\tau^{ei}_{\nu}} + \\
\Omega_{do}(1-e^{-\tau^{do}_{\nu}})B_{\nu}(T_{do})e^{-\tau^{ei}_{\nu}} + \\
\Omega_{ei}(1-e^{-\tau^{ei}_\nu})B_{\nu}(T_{ei}) + \\
\Omega_{ff}(1-e^{-\tau^{ff}_\nu})B_{\nu}(T_{ff})e^{-\tau^{di}_{\nu}-\tau^{do}_{\nu}-\tau^{ei}_{\nu}}, 
\end{split}
\end{equation}
where $\Omega_{ff,di,do,ei}$ are the solid angles of the free-free emission source, the inner disk, outer disk, and inner envelope components (assuming $\Omega_{ei}$$>$$\Omega_{do}$$>$$\Omega_{di}$$>$$\Omega_{ff}$); $\tau^{ff,di,do,ei}_{\nu}$ and $T_{ff,di,do,ei}$ are the optical depths and dust temperatures of these components.
Following \citet{Keto2003} and \citet{Mezger1967}, we approximate $\tau_{\nu}^{ff}$ by:
\begin{equation}
\label{eq:tauff}
\begin{split}
\tau_{\nu}^{ff}=  \\
8.235\times10^{-2}\left(\frac{T_{e}}{\mbox{K}}\right)^{-1.35}\left(\frac{\nu}{\mbox{GHz}}\right)^{-2.1}\left(\frac{\mbox{EM}}{\mbox{pc\,cm$^{-6}$}}\right), \\
\end{split}
\end{equation}
where $T_{e}$ is the electron temperature, and EM is the emission measure defined as EM$=$$\int n_{e}^{2}d\ell$, {with $n_{e}$ being the electron number density}.
Here we assume that the free-free emission source is obscured by all dust components.
The following discussion will not be sensitive to whether or not the free-free emission source is obscured.

\begin{figure}
\hspace{-0.5cm}
\begin{tabular} { p{4cm} p{4cm} }
\includegraphics[width=4.7cm]{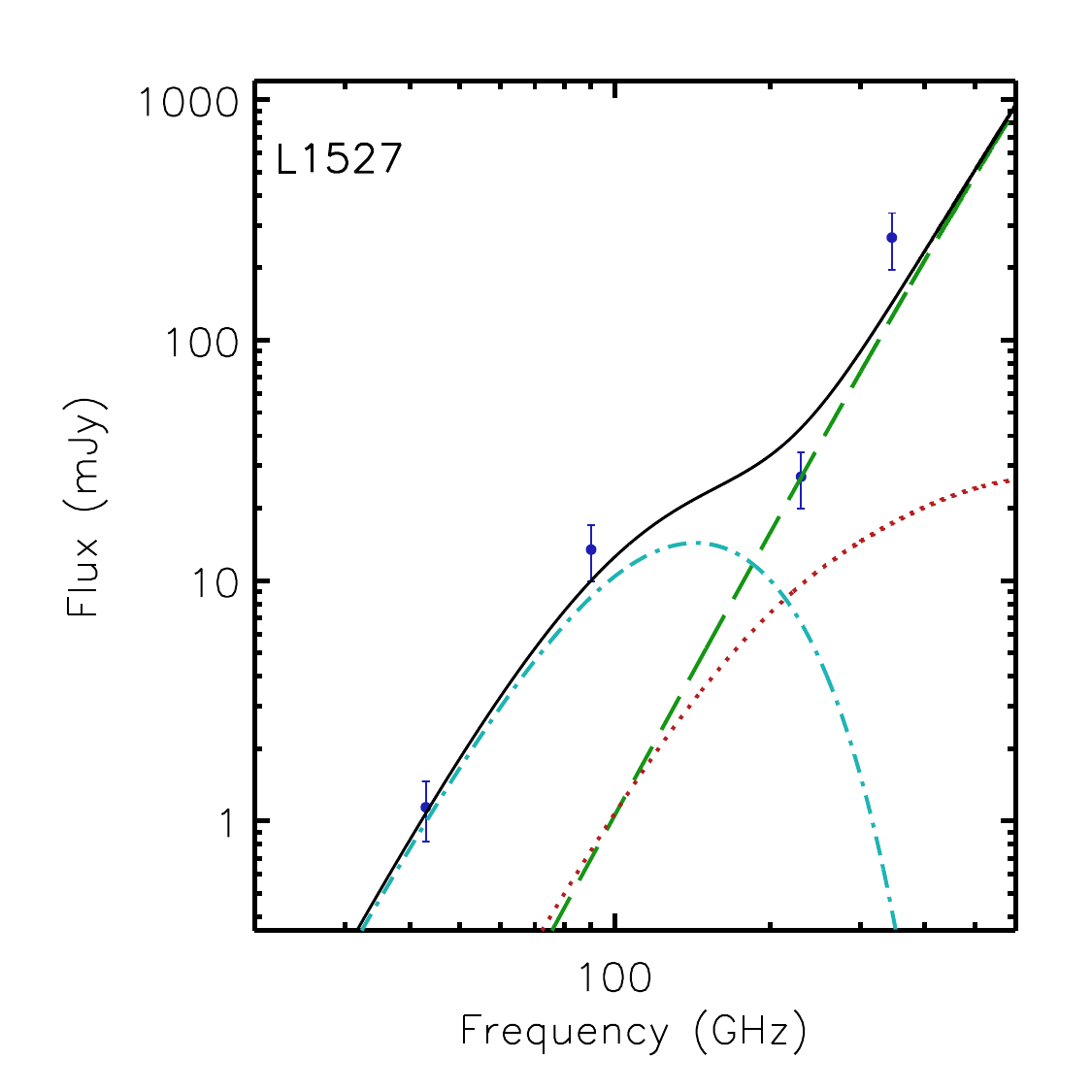} &
\includegraphics[width=4.7cm]{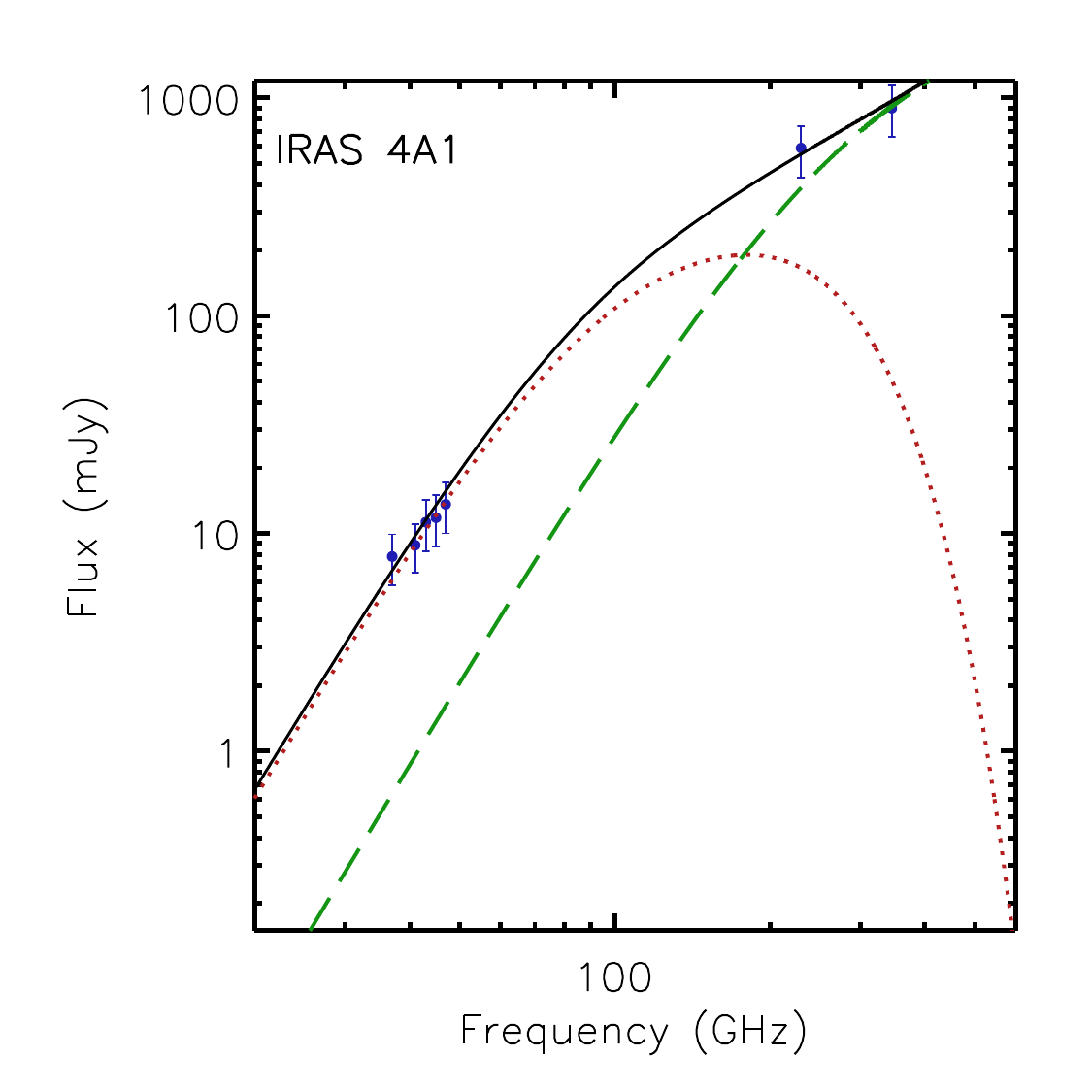} \\
\includegraphics[width=4.7cm]{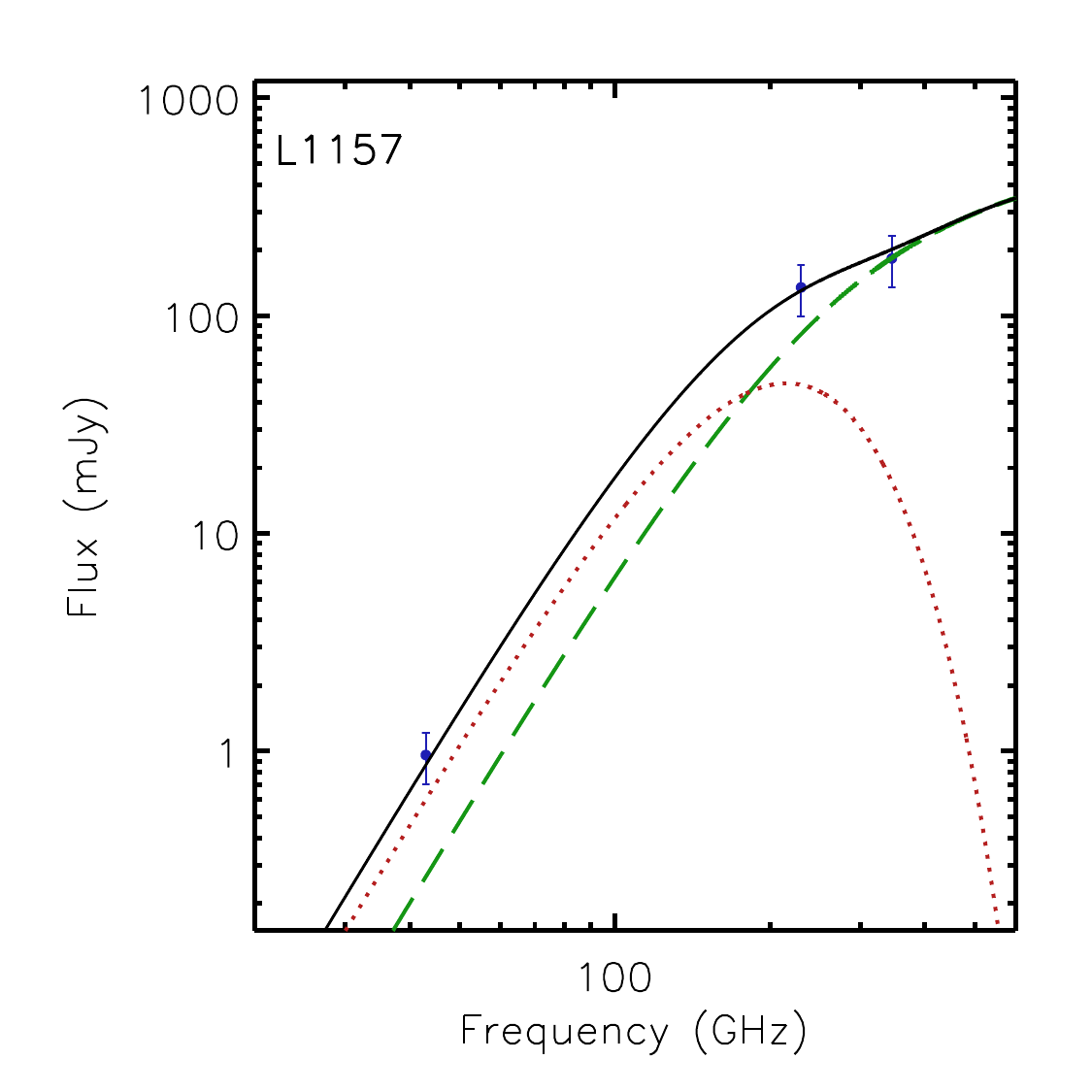} &
\includegraphics[width=4.7cm]{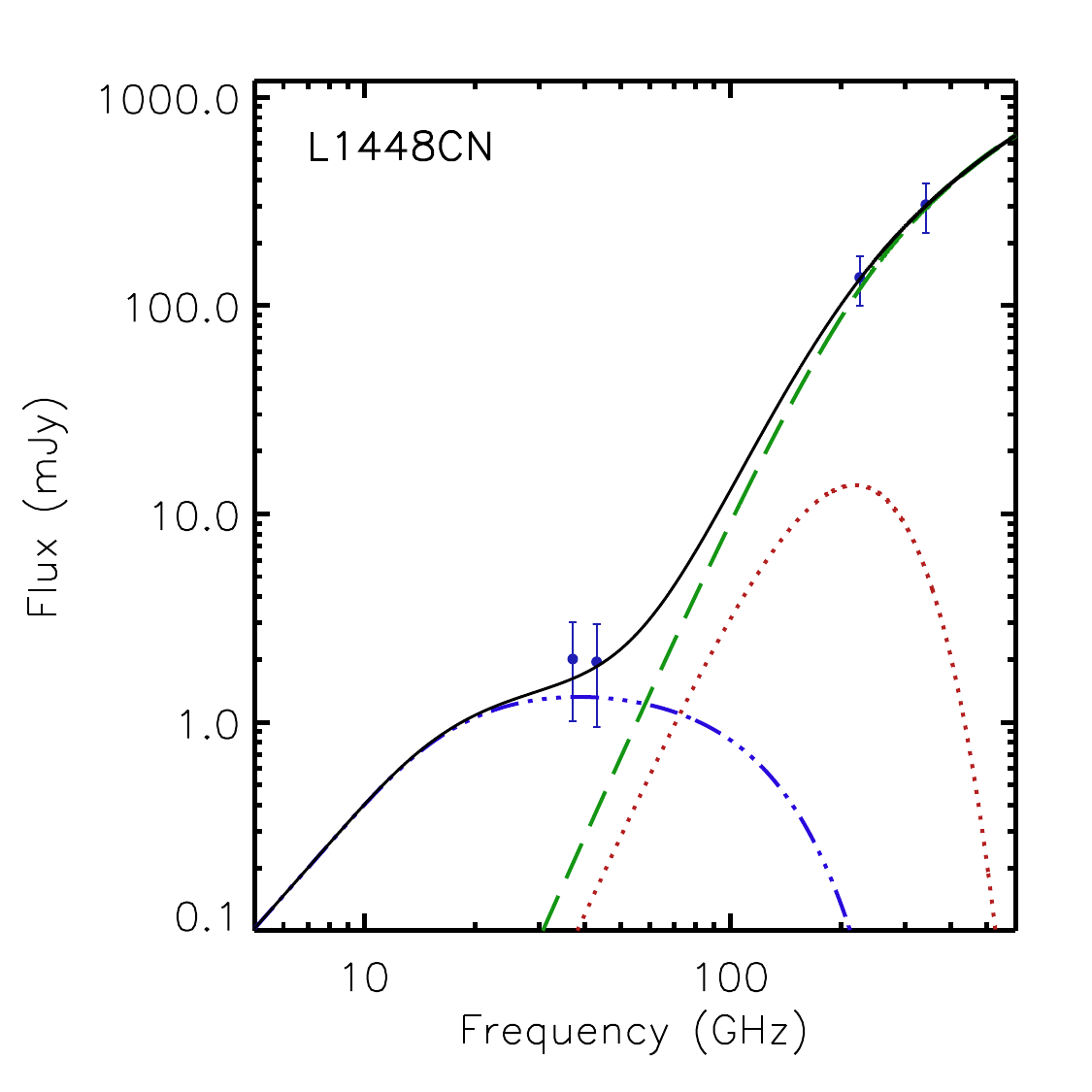} \\
\includegraphics[width=4.7cm]{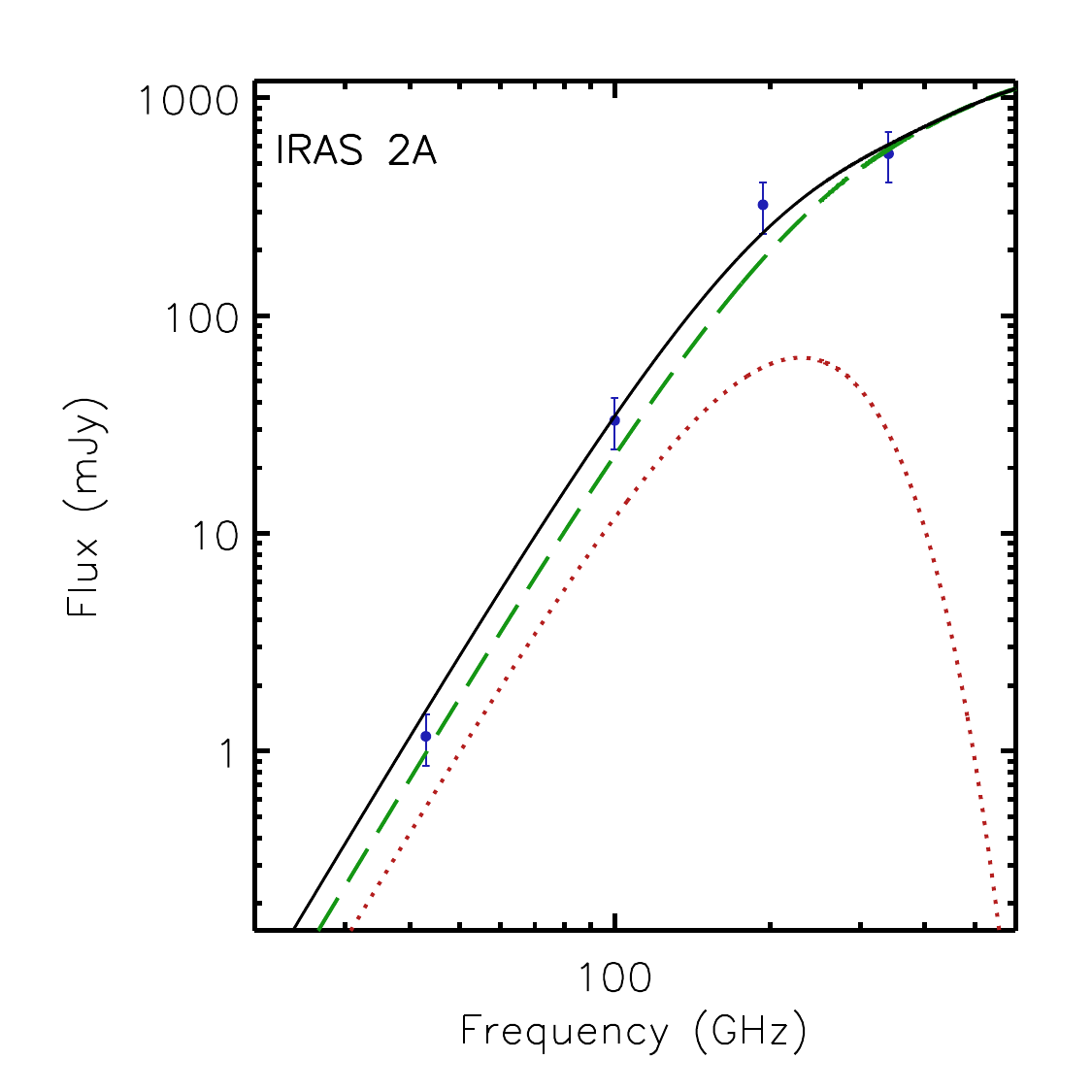} & \\
\end{tabular}
\caption{Toy SED models for sources with spectral indices $\alpha$$<$3.5. In these models, we assume $\beta$$=2.0$ for all components. Dashed-dotted (cyan), dotted (red), dashed (green), and dashed-triple-dotted (blue) lines show the contribution of (sub)millimeter fluxes from the inner disk components, the outer disk components, the inner envelope components, and the free-free emission, respectively. Solid lines show integrated fluxes from all components.}
\label{fig:toymodel}
\vspace{0.4cm}
\end{figure}

{Our goal is to test if (a) the SEDs of L1448CN, IRAS41, L1157, L1527 can be explained without having to assume a significant affectation by dust grain growth, and (b) if the $\alpha<$2.0 spectral index of IRAS2A at 230--345 GHz can be explained in terms of obscuration. We construct toy SED models for all these sources based on Equation \ref{eqn:multicomponent}  and assuming $\beta$=2.0. The parameters we used are listed in Table \ref{tab:3components} and the results are presented in Figure \ref{fig:toymodel}. Notice that the number of free parameters in our toy model is larger than the number of measurements we have. Therefore, it is impossible to avoid a degeneracy in the product of the dust opacity at 230 GHz and column density, $\kappa_{\mbox{\tiny{230 GHz}}}$$\Sigma$
 and the dust temperature, T.}
Nevertheless, we argue that these toy models only use the minimum numbers of dust or free-free emission components and thus the minimum numbers of free parameters that are required to explain the observed SED.
For example, two dust emission components are required to reproduce the $\alpha$$<$2.0 results of {L1527 between} 90 GHz and 230 GHz, and yet another optically thinner dust component {is needed to} reproduce its high flux at 345 GHz.
For this particular case, the SED peak of the hot inner disk component (i.e., Figure \ref{fig:toymodel}, {dash-dot cyan curve}) is shifted to a lower frequency than what is expected from {a} single emission component, modified black body profile (Equation \ref{eq:F_nu}, \ref{eq:taudust}) due to the obscuration by other components.
For L1448CN, L1157, IRAS2A, and IRAS4A1, a cool inner envelope component is required to obscure an embedded warm disk component to reproduce the $\alpha$$<$2.0 results in between 230 GHz and 345 GHz.
Similarly, the SED peaks of the outer disk components of these four sources (i.e., Figure \ref{fig:toymodel}, {red dotted curve}) are shifted to the lower frequencies, due to the obscuration by the envelope components.
The SEDs of these four sources can be modeled without considering the innermost hot disk component.
The contribution of the innermost hot disk component is uncertain for L1448CN due to the confusion by free-free emission.
We {thus assume that the hot inner disk component is} negligible in our present toy models.
We caution that in order to minimize the number of free parameters, our toy models may be yet over simplified.
In reality, dust temperature may have rather continuous spatial distributions.

We find that our simplified toy models can reasonably reproduce the observed SEDs of all five sources.
This implies that the actual $\beta$ may be consistent with 2.0, which is higher than the fitted values presented in Figure \ref{fig:sed}, \ref{fig:2beta}, and Table \ref{tab:results}.
Assuming a gas-to-dust mass ratio of 100.0, the implied gas mass of individual dust components can be approximated by
\begin{equation}
\left(\frac{m_{gas}}{M_{\odot}}\right)=4.8\times10{^5}\left(\frac{   \Omega\times\kappa_{\mbox{\tiny{230GHz}}}\Sigma   }{ \kappa_{\mbox{\tiny{230GHz}}}  }\right)\times
\left(\frac{ \mbox{\scriptsize{distance}}  }{ \mbox{\small{pc}}  }\right)^2 .
\end{equation}
The estimated integrated gas masses for L1527, IRAS\,4A1, L1157, L1448CN, and IRAS2A are 0.084/$\kappa_{\mbox{\tiny{230GHz}}}$$M_{\odot}$, 5.2/$\kappa_{\mbox{\tiny{230GHz}}}$$M_{\odot}$, 0.99/$\kappa_{\mbox{\tiny{230GHz}}}$$M_{\odot}$, 0.34/$\kappa_{\mbox{\tiny{230GHz}}}$$M_{\odot}$, and 1.0/$\kappa_{\mbox{\tiny{230GHz}}}$$M_{\odot}$, respectively.
Assuming $\kappa_{\mbox{\tiny{230GHz}}}$$\sim$1, the estimated gas masses are reasonable for young, low-mass star-forming sources.

\section{Discussion}\label{sec:discussion}
\subsection{{Alternative explanation for low $\beta$}} \label{sec:discussion2}

If the simplified geometric picture and temperature profile of our toy SED models are reasonable, then the previously reported, radially increasing $\alpha$ \citep{chiang2012envelope} may be alternatively interpreted by that the embedded hot dust component progressively becomes more prominent towards the projected central region (i.e., becoming warmer or denser), instead of by radially increasing $\beta$.
On the other hand, perhaps not all the inner dust components are fully obscured by the outer ones.
When dust emission from the inner region is not obscured (e.g., when the disk components are approximately in face-on projection), the sparsely sampled SED will look similar to a single modified black body profile unless the temperature changes drastically.

Comparing the toy models presented in Figure \ref{fig:toymodel} with the systematic fittings presented in Figure \ref{fig:sed} leads us to think that when the SEDs cannot be well represented by a single modified emission profile (Equation \ref{eq:F_nu}), the full single-component modified black body fittings does not provide a more physical derivation of $\beta$ than the derivation based on the optically thin and Rayleigh-Jeans assumption.
Instead, they may converge to equally poor (or good) fits.
One may face the situation that the assumption of single dominant emission component breaks down when a too wide frequency range is considered, and that the high frequency emission and the low frequency emission do not necessarily trace the same emission component (e.g., Figure \ref{fig:toymodel}).
The $\beta$ value of the dominant contributor for low frequency (e.g., $\lesssim$50 GHz) emission may be more accurately constrained when there are two independent measurements from the optically thinner part of the spectrum, which may be (close to) the case of IRAS\,4A2 reported by \citet{cox2015VANDAM}.

\subsection{Implications for dust grain growth in Class 0 YSOs}

\begin{figure}
\centering
\includegraphics[width=8cm]{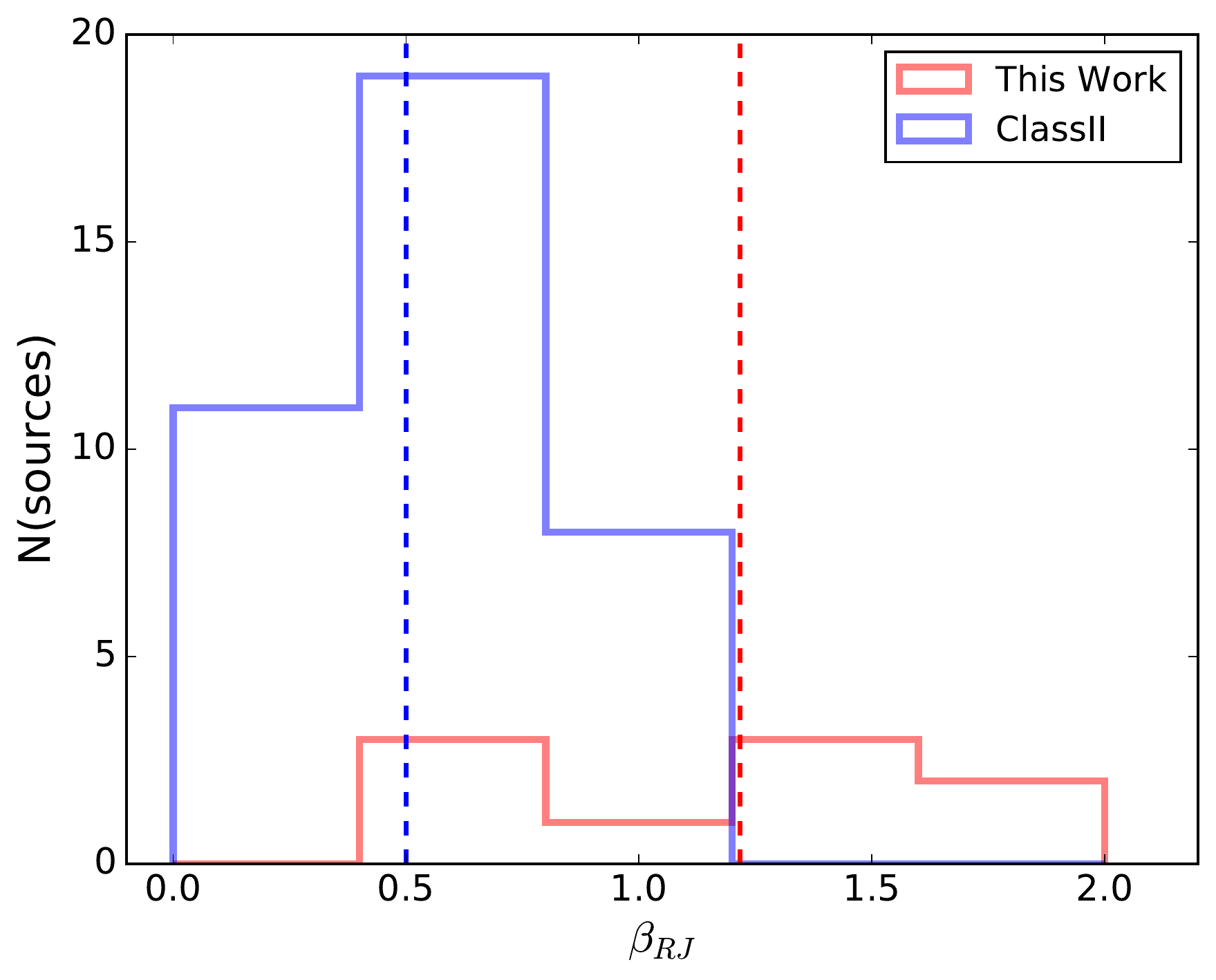}
\caption{Distribution of $\beta_{RJ}$ in different evolutionary stages. 
Class II sample includes all sources from Ricci 2010a, 2010b. Dashed lines indicate the median of each sample.}
\vspace{0.4cm}
\label{fig:hist}
\end{figure}

We are now in a position to discuss implications of our results that are derived from the SED analysis for our targets.

Our SED fitting with the single emission component indicates that the resultant value of $\beta$ becomes comparable to the value of the interstellar dust for five of the nine targets (IRAS 2A, IRAS4A2, B1bN, B1bS, and SerFIRS1, see Figure \ref{fig:2beta}). 
This suggests that dust grain growth is unlikely to proceed significantly in these targets. 
For the remaining four sources (L1448CN, NGC\,1333\,IRAS4A1, L1157, L1527), the single component SED fitting shows a lower ($<1$) value of $\beta$, which can be attributed to a higher degree of dust growth.
{However, by invoking a multiple emission component scenario, we can account for} the observed SEDs reasonably well under the condition that $\beta=2$. 
Thus, we argue that the values of $\beta$ and $\beta_{RJ}$ for these four targets should be regarded as lower limits. 
As a consequence, our presented {SED analysis does not provide firm evidence of significant dust grain growth to reduce the value of $\beta$ down to an order of unity or less in our nine} Class 0 YSO targets.
Note that the additional information of the target sources, especially time variability,
would be valuable to examine how our interpretations can be affected by such effects.
All kinds of pure radiative transfer simulations/modeling (including our model) 
generally assume equilibrium states for gas and dust in the targets.

How is our interpretation compatible with the previous studies?
A lower degree of dust growth in Class 0 YSOs is indeed in favor for the results of theoretical studies. 
Some studies show that the collisional growth of dust grains generally does not proceed up to the grain size of $\sim 1$mm due to the low density, short-lived environment there \citep[e.g.,][]{Ormel2009,Hirashita2013}. 
Even if some degrees of grain growth can occur, the number density of larger grains may be low \citep[e.g.,][]{Wong2016} and/or the resultant dust particles would be porous aggregates \citep[e.g.,][]{Ormel2007, Okuzumi2012}, 
so that a signature of grain growth may not be captured well by our SED fitting. 
Thus, our interpretation can hold in the current literature.

The above interpretation {enables in turn, to imply that dust grain growth may not become significant enough to lower the value of $\beta$ in a stage earlier than late-Class 0}. 
{On the other hand, multiple studies have reported that the values of $\beta$ already go down to an order of unity or less at the Class II stage.
Comprehensive surveys \citep{ricci2010a, ricci2010b} on 38 flux selected class II sources in the $\rho$-Ophiuchi and Taurus-Auriga star forming region (SFR) revealed $\langle\beta\rangle$=0.5--0.6.
They fitted the SEDs of resolved sources at the wavelengths of 0.5--3mm with a radiative transfer model of two-layers disks, proposed by \cite{chiang1997spectral}, and concluded that maximum grain size $a_{max}$ $>$ 1cm is necessary to explain the derived $\beta$ in both SFR considering a reasonable range of dust opacity and power-law size distribution \citep[see Figure 3 of][]{ricci2010a}.}
The distribution of $\beta_{RJ}$ of their Class II sources and our sources are compared in Figure \ref{fig:hist}.
Taking into account these results, it may be crucial to perform a survey for Class I YSOs to detect the evolutionary history of $\beta$ due to grain growth.

Finally, Figure \ref{fig:hist} shows a relative uniform distribution of the $\beta$ derived from our systematic analysis of the SED fitting for our Class 0 smaple.
If this flat distribution would arise from the effect of the multiple emission components, then this trend may imply that our Class 0 YSOs harbor randomly orientated disk-like structures, such that the hot inner parts of only some of them are obscured (at shorter than millimeter wavelength bands), or that the heating mechanism(s) for the innermost disk is episodic (e.g. \citealt{Vorobyov2013}). These hypotheses may be tested by observing a statistical sample of Class 0 and Class I objects, and by exploring their gas velocity profiles.

\section{Conclusions}\label{sec:conclusion}
We have analyzed the archival SMA, (J)VLA, NMA, CARMA, and ALMA data for nine well studied Class 0 YSOs, which cover $\sim$30 GHz to $\sim$345 GHz frequency ranges.
We systematically perform single-component modified black body fittings to their SEDs at radio and (sub)millimeter wavelengths, and find that five of them show a dust opacity spectral index $\beta$$\sim$1.7, which implies no significant signatures of dust grain growth.

We cannot rule out that dust grain growth has significantly lowered the values of $\beta$ in the remaining four sources.
However, our radiative transfer toy models assuming $\beta$$=$2.0 show that their SEDs can alternatively be interpreted by the presence of hot inner dust components that are obscured by the cooler and optically thicker outer components at short wavelengths (e.g., $<$1 mm).
In fact, this mechanism is favored to explain the {spectral slope (alpha) values lower than 2.0} observed in certain frequency ranges of some sources.
Therefore, the presented data in this work do not show robust evidence that a higher degree of dust grain growth has taken place in any of our observed sources.
By comparing with the previous observations of Class II sources, we suggest that dust grain growth may significantly reduce the values of $\beta$ in circumstellar {(pseudo-)}disks or envelopes no earlier than the late-Class 0 stage but before the Class II stage of YSOs.
A future survey of radio and (sub)millimeter SEDs for Class I YSOs may improve the constraints on the timescales of grain growth and the starting point where a lower ($\la 1$) value of $\beta$ is achieved, and thereby improving our understanding of planet-forming environments and the formation of planetesimals.

\begin{acknowledgements}
The authors thank the anonymous referee for useful comments on our manuscript.
The Submillimeter Array is a joint project between the Smithsonian
Astrophysical Observatory and the Academia Sinica Institute of Astronomy
and Astrophysics, and is funded by the Smithsonian Institution and the
Academia Sinica \citep{Ho2004}.
The National Radio Astronomy Observatory is a facility of the National
Science Foundation operated under cooperative agreement by Associated
Universities, Inc.
{This paper makes use of the following ALMA data: ADS/JAO.ALMA\#2011.0.00210.S. 
ALMA is a partnership of ESO (representing its member states), NSF (USA) and NINS (Japan), together with NRC (Canada), NSC and ASIAA (Taiwan), and KASI (Republic of Korea), in cooperation with the Republic of Chile. The Joint ALMA Observatory is operated by ESO, AUI/NRAO and NAOJ.}
Part of this research was carried out at Jet Propulsion Laboratory, California Institute of Technology, 
under a contract with NASA. 
Y.H. is supported by JPL/Caltech.
N.H. is supported by the grant from Ministry of Science and Technology of Taiwan (MoST 105-2112-M-001-026).
\end{acknowledgements}

\renewcommand
\refname{Reference}
\bibliographystyle{apj}
\bibliography{ref}

\end{document}